\documentclass[fleqn,usenatbib]{mnras}

\usepackage{newtxtext,newtxmath}

\usepackage[T1]{fontenc}

\DeclareRobustCommand{\VAN}[3]{#2}
\let\VANthebibliography\thebibliography
\def\thebibliography{\DeclareRobustCommand{\VAN}[3]{##3}\VANthebibliography}

\usepackage{graphicx}	
\usepackage{amsmath}	
\usepackage{fdsymbol}

\title[Velocity-dependent SIDM haloes]{Cosmological and idealized simulations of dark matter haloes with velocity-dependent, rare and frequent self-interactions}

\author[M. S. Fischer et al.]{Moritz S. Fischer,$^{1,2,3}$\thanks{E-mail: mfischer@usm.lmu.de (LMU)}
Lenard Kasselmann,$^{3}$
Marcus Br\"{u}ggen,$^{3}$
Klaus Dolag,$^{1,4}$
\newauthor{Felix Kahlhoefer,$^{5}$
Antonio Ragagnin,$^{6,7,8}$
Andrew Robertson,$^{9}$}
Kai Schmidt-Hoberg$^{10}$
\\
$^{1}$Universit\"ats-Sternwarte, Fakult\"at für Physik, Ludwig-Maximilians-Universit\"at M\"unchen, Scheinerstr. 1, D-81679 M\"unchen, Germany\\
$^{2}$Excellence Cluster ORIGINS, Boltzmannstrasse 2, D-85748 Garching, Germany\\
$^{3}$Hamburger Sternwarte, Universit\"at Hamburg, Gojenbergsweg 112, D-21029 Hamburg, Germany\\
$^{4}$Max-Planck-Institut f\"ur Astrophysik, Karl-Schwarzschild-Str. 1, D-85748 Garching, Germany\\
$^{5}$Institute for Theoretical Particle Physics (TTP), Karlsruhe Institute of Technology (KIT), D-76128 Karlsruhe, Germany\\
$^{6}$Dipartimento di Fisica e Astronomia "Augusto Righi", Alma Mater Studiorum Università di Bologna, via Gobetti 93/2, I-40129 Bologna, Italy\\ 
$^{7}$INAF-Osservatorio Astronomico di Trieste, via G. B. Tiepolo 11, I-34143 Trieste, Italy\\
$^{8}$IFPU -- Institute for Fundamental Physics of the Universe, Via Beirut 2, I-34014 Trieste, Italy\\
$^{9}$Jet Propulsion Laboratory, California Institute of Technology, 4800 Oak Grove Drive, Pasadena, CA 91109, USA\\
$^{10}$Deutsches  Elektronen-Synchrotron  DESY,  Notkestr.~85, D-22607  Hamburg,  Germany\\
}

\date{Accepted XXX. Received YYY; in original form ZZZ}

\pubyear{2023}

\begin{document}
\label{firstpage}
\pagerange{\pageref{firstpage}--\pageref{lastpage}}
\maketitle

\begin{abstract}

Dark matter self-interactions may have the capability to solve or at least mitigate small-scale problems of the cosmological standard model, Lambda cold dark matter.
There are a variety of self-interacting dark matter (SIDM) models that lead to distinguishable astrophysical predictions and hence varying success in explaining observations.
Studies of dark matter (DM) density cores on various mass scales suggest a velocity-dependent scattering cross-section.
In this work we investigate how a velocity dependence alters the evolution of the DM distribution for \emph{frequent} DM scatterings and compare to the velocity-independent case. 
We demonstrate that these cases are qualitatively different using a test problem.
Moreover, we study the evolution of the density profile of idealized DM haloes and find that a velocity dependence can lead to larger core sizes and different time-scales of core formation and core collapse.
In cosmological simulations, we investigate the effect of velocity-dependent self-interaction on haloes and satellites in the mass range of $\approx 10^{11}$--$10^{14} \, \mathrm{M_\odot}$.
We study the abundance of satellites, density, and shape profiles and try to infer qualitative differences between velocity-dependent and velocity-independent scatterings as well as between frequent and rare self-interactions.
We find that a strongly velocity-dependent cross-section can significantly amplify the diversity of rotation curves, independent of the angular dependence of the differential cross-section. We further find that the abundance of satellites in general depends on both the velocity dependence and the scattering angle, although the latter is less important for strongly velocity-dependent cross-sections.
\end{abstract}




\section{Introduction}

Historically, dark matter (DM) self-interactions have been motivated to solve problems on small, i.e.\ galactic scales.
It was found that cosmological DM-only simulations can explain the large-scale structure of the universe quite well.
But on smaller scales, deviations between the observations and simulations were encountered \citep[e.g.][]{Moore_1998}.
\cite{Spergel_2000} proposed self-interacting dark matter (SIDM) as a solution to two problems on small scales.
Namely, SIDM can reduce the abundance of satellites and the central density of haloes.
As the self-interactions lead to heat flow into the central region of a Navarro--Frenk--White \citep[NFW;][]{Navarro_1996} halo, they reduce the central density and can form density cores.
The first $N$-body simulation using a Monte Carlo scheme of this core formation has been performed by \cite{Burkert_2000}.
Since then SIDM has been found to be capable of solving or at least mitigating further small-scale problems of cold dark matter \citep[CDM; for a review see][]{Tulin_2018, Adhikari_2022}.
This does not only include the core-cusp problem \citep[e.g.][]{Dave_2001}, but also diverse rotation curves \citep[e.g.][]{Creasey_2017, Kamada_2017, Robertson_2018, Correa_2022} and the too-big-to-fail problem \citep[e.g.][]{Zavala_2013, Elbert_2015, Kaplinghat_2019a}.
For a review of small-scale problems in Lambda cold dark matter ($\Lambda$CDM), we refer the reader to \cite{Bullock_2017}.

Meanwhile, it has also emerged that there are other avenues to solve these small-scale problems.
On the one hand, it was found that including the baryonic physics, in particular, feedback mechanisms from supernovae \citep[e.g.][]{Read_2005, Governato_2012,Pontzen_2012} and black holes can form density cores \citep[e.g.][]{Martizzi_2013, Silk_2017, Peirani_2017}.
On the other hand, researchers have become more cautious about inferring density profiles from rotation curves \citep[e.g.][]{Pineda_2016, Read_2016b, Genina_2018, Oman_2019, Roper_2023, Downing_2023}.
Beyond SIDM, other DM models have been investigated, including warm DM \citep{Dodelson_1994} and fuzzy DM \citep{Hu_2000}.

Although SIDM has initially been mainly motivated by small-scale issues, it provides DM candidates worth investigating, independent of the state of the small-scale crisis.
The nature of DM is still unknown and could have properties which we can only infer indirectly via astronomical observations. This is true for models of SIDM, and studying them is essentially constraining particle physics properties of DM.
Particle candidates that fall into the class of SIDM can have various characteristics. The scattering may be elastic or inelastic, it may involve multiple states and can feature different angular dependencies.
Another aspect is how the cross-section depends on the relative velocity of the scattering particles.

Velocity-dependent self-interactions have been recently studied by various authors \citep[e.g.][]{Colin_2002, Nadler_2020, Yang_2022D, Outmezguine_2023, Yang_2023S}.
Such studies were performed not only with DM-only (DMO) simulations but also within hydrodynamical cosmological simulations \citep[e.g.][]{Vogelsberger_2014, Robertson_2019, Robertson_2020, Rose_2022, Mastromarino_2023, Rahimi_2023}.
They are well-motivated for different angular dependencies, including forward-enhanced cross-sections from light mediator models \citep[e.g.][]{Buckley_2010, Loeb_2011, Bringmann_2017}.
But also models of resonant scattering \citep[e.g.][]{Chu_2019, Tsai_2022} can explain a velocity dependence while featuring an isotropic cross-section.

From an astronomical perspective, velocity-dependent self-interactions are well motivated \citep[e.g.][]{Kaplinghat_2016, Correa_2021, Gilman_2021, Sagunski_2021, Silverman_2022, Lovell_2023}.
They would allow fulfilling stringent constraints from galaxy clusters while having a fairly large effect on low-mass haloes.
When the self-interaction cross-section decreases with velocity, it has a weaker effect in galaxy clusters because their typical relative DM velocities are larger than in galaxies.
Furthermore, they can lead to a qualitative different evolution of systems that involve multiple velocity scales.
For instance, this is true for the evolution of the satellite distribution \citep[e.g.][]{Zeng_2022} and could lead to an increase in the diversity of density profiles and rotation curves \citep[e.g.][]{Nadler_2023, Yang_2023Da}.

The aim of this study is to explore qualitative differences arising from the velocity dependence of the self-interactions and to understand their implications on constraining the angular dependence of the cross-section.
In this paper, we consider two different angular dependencies:
First, isotropic scattering, to which we refer as rare self-interactions (rSIDM).
Secondly, a cross-section with typical scattering angles that are very small.
In consequence, frequent interactions are needed to significantly alter the DM distribution.
Hence, we call it frequent self-interactions (fSIDM).

In contrast to previous studies of anisotropic cross-sections \citep[e.g.][]{Robertson_2017b, Banerjee_2020, Correa_2022, Yang_2022D}, we study a limit where the momentum transfer is kept constant, but the typical scattering angle is approaching zero, while the scattering rate increases.

Frequent self-interactions show a drag-like behaviour \citep{Kahlhoefer_2014} and are known for being capable of producing large offsets between the galaxies and the DM component in merging galaxy clusters \citep[e.g.][]{Fischer_2021a, Fischer_2023b}.
In addition, it has been found that they are more efficient in suppressing the abundance of satellites compared to an isotropic cross-section \cite{Fischer_2022} and may alter the morphology of satellite galaxies \citep{Secco_2018, Pardo_2019}.
These signatures could potentially allow to constrain the angular dependence of DM self-interactions.
However, fSIDM is mainly motivated by light mediator models which have velocity-dependent cross-sections.
But the aforementioned results are from studies of velocity-independent models.
In consequence, it is crucial to extend them to models featuring a velocity dependence – an aim of this paper.

We explore rSIDM and fSIDM models with several velocity dependencies to study qualitative differences arising from the velocity and angular dependence.
The scattering of all SIDM models we consider is elastic.
For our study, we employ idealized \textit{N}-body simulations of a test problem and DM haloes as well as cosmological simulations.
Unlike velocity-independent models \citep{Fischer_2022}, fSIDM with a velocity-dependent interaction has not been studied in a cosmological context.
Finally, all our simulations are DM-only, i.e.\ we ignore the effects of baryons.
In a companion paper \citep[][]{Sabarish_2023}, velocity-dependent fSIDM is studied in the context of merging galaxy clusters.

This paper is structured as follows.
In Section~\ref{sec:numerical_setup}, we describe the numerical set-up of our simulations including a novel time-stepping criterion.
A presentation of the simulations and our results follows for the idealized set-ups in Section~\ref{sec:idealised_sims} and the cosmological simulations in Section~\ref{sec:cosmo_sims}.
Shortcomings and directions for further research are discussed in Section~\ref{sec:discussion}.
Finally, in Section~\ref{sec:conclusions} we conclude.
Additional information can be found in the appendices.


\section{Numerical set-up} \label{sec:numerical_setup}

In this section, we describe our numerical set-up.
First, we begin by describing the simulation code and the SIDM implementation.
We continue with the parametrization for the velocity-dependent cross-section.
Next, we introduce a novel time-step criterion for the velocity-dependent self-interaction.
Lastly, the simulations with their initial conditions and the identification of the substructure are described.
In addition, a description of our improved parallelization scheme for SIDM can be found in Appendix~\ref{sec:parallelisation}.

\subsection{SIDM implementation and simulations}  \label{sec:simulations}

For our simulations, we use the cosmological hydrodynamical $N$-body code \textsc{opengadget3}.
The predecessor \textsc{gadget-2} has been described in \cite{Springel_2005}.
Various additional modules have been developed for the \textsc{opengadget3} version that we are using.
The implementation of DM self-interactions has been described by \cite{Fischer_2021a, Fischer_2021b, Fischer_2022}.

The SIDM module in \textsc{opengadget3} is capable of modelling very anisotropic cross-sections.
Precisely speaking, we model the limit where the momentum transfer is kept constant, but the size of the scattering angles is approaching zero.
In this limit, the number of scattering events becomes infinite, which is why we call it frequent self-interactions.
For very anisotropic cross-sections the self-interactions can be effectively described as a drag force \citep{Kahlhoefer_2014}.
The numerical scheme computes the interactions between the numerical particles in a pairwise manner.
We use the drag force and apply it to each pair of close particles to model the frequent self-interactions.
To conserve energy, we add momentum in a random direction but perpendicular to the direction of motion of the particles for each pair.
In consequence, our scheme is a Monte-Carlo scheme like other state-of-the-art schemes for SIDM.
The fSIDM scheme models only the limit of an extremely anisotropic cross-section and cannot reproduce arbitrary angular dependencies.
To date, this is the only implementation for simulating fSIDM.

The code is also able to simulate isotropic cross-sections.
Given that the scattering rate of physical particles is very infrequent for momentum transfer cross-sections allowed in astrophysical systems, we refer to it as rare scattering. 
Interactions between numerical particles are modelled in a pairwise manner too.
For close particles, an interaction probability is computed and by drawing a random number one decides whether two particles interact. 
Given that they interact, they are treated analogously to physical particles scattering about each other.
The employed scheme \citep[described by][]{Fischer_2021a} is very similar to the one introduced by \cite{Rocha_2013}, except that we use an adaptive kernel size set by the 64 next neighbours and a different time-step criterion.
Another advantage of our SIDM module is that it conserves energy explicitly. Energy non-conservation typically arises when a numerical particle scatters at the same time with multiple partners using the same velocity. Avoiding this is particularly challenging for parallel computations. An alternative to our approach to overcome this problem has been recently presented by \cite{Valdarnini_2023}.

We have run several simulations of CDM, rSIDM, and fSIDM for idealized set-ups with individual haloes as well as cosmological simulations.
For all simulations, we used the cosmological $N$-body code \textsc{opengadget3}.
The details of the simulations can be found in the corresponding Sections~\ref{sec:idealised_sims} and \ref{sec:cosmo_sims}.
In addition, we ran simulations to test the code, they can be found in the Appendices~\ref{sec:comoving_integration_test} and \ref{sec:convergence_test}.

\subsection{Velocity-dependent cross-section}

There are  numerous studies in the literature considering a cross-section, $\sigma$, that depends on the scattering velocity, $v$. A typical choice – that we employ as well – is a cross-section that scales as $\sigma \propto v^{-4}$ in the limit of high $v$. This dependence may be motivated by particle physics \citep[e.g.][]{Ibe_2010, Tulin_2013a} and has been employed in numerous studies \citep[e.g.][]{Kaplinghat_2016, Robertson_2017b}. 

Following \cite{Kahlhoefer_2017} and \cite{Robertson_2017b}, we consider the momentum transfer cross-section
\begin{equation} \label{eq:momentum_transfer_cross_section}
\sigma_\mathrm{T}=2 \uppi \int_{-1}^{1} \frac{\mathrm{d} \sigma}{\mathrm{d} \Omega_{\mathrm{cms}}}\left(1-|\cos \theta_{\mathrm{cms}}| \right) \mathrm{d} \cos \theta_{\mathrm{cms}}
\, .
\end{equation}
We parameterize the velocity dependence of the momentum transfer cross-section as
\begin{equation} \label{eq:veldep}
    \frac{\sigma_\mathrm{T}}{m} = \frac{\sigma_0}{m} \left(1+ \left( \frac{v}{w} \right)^\beta \right)^{\alpha/\beta} \,.
\end{equation}
Here, $\sigma_0$ corresponds to the cross-section in the velocity-independent regime, $w$ denotes the velocity cutoff, $\alpha$ sets the decline at high velocities and $\beta$ describes the transition from the constant cross-section at low velocities to the decreasing cross-section at high velocities.
In this study, we always set $\alpha = -4$ and $\beta = 2$.
This choice is motivated by the fact that in the limit of the Born-approximation, the velocity dependence of the total and the transfer cross-section are very similar \citep{Ibe_2010}.
More details on the transfer cross-section and the possible connections to the underlying particle physics can be found in the companion paper \citep[][]{Sabarish_2023}.

In most physically motivated cases, a velocity dependence also implies an angular dependence of the differential scattering cross-section. 
$N$-body simulations had been limited in simulating frequent scatterings about small angles until the work by \cite{Fischer_2021a}. Here, we go beyond the common large-angle scattering and investigate small-angle as well as isotropic scattering combined with a velocity
dependence.

In order to probe different velocity regimes, we use several combinations of $\sigma_0$ and $w$.
These are described together with the details of the simulations in Sections~\ref{sec:idealised_sims} and~\ref{sec:cosmo_sims}.
Each parameter set is simulated with fSIDM and rSIDM, the latter corresponding to isotropic scattering.
Note that we use the momentum transfer cross-section (equation~\ref{eq:momentum_transfer_cross_section}) to match fSIDM and rSIDM.
In the case of isotropic scattering, the total cross-section is twice as large as the momentum transfer cross-section.

\subsection{Time-step criterion} \label{sec:time_step_crit}

For velocity-dependent self-interactions, a separate time-step criterion can become more important than for velocity-independent scatterings because cross-sections can become large at low velocities.
Depending on the cross-section this can give more stringent limitations on the time-step than imposed by the gravity scheme.
We found that the time-step criterion introduced by \cite{Fischer_2021b} for velocity-independent self-interactions is not always well-suited for a velocity-dependent cross-section \citep[this has been previously described by][]{Kasselmann_2021}.
The difficulty arises from estimating the scattering velocity for which the effect from the self-interactions is strongest and thus requires the smallest time-step.
Concerning the value range of scattering velocities a particle may see.
For a velocity-independent cross-section, this is simply the maximal scattering velocity.
But for a velocity-dependent cross-section, it is typically smaller and the estimate using the criterion of \cite{Fischer_2021b} would be more noisy and unnecessarily complicated.

Here, we introduce a new time-step criterion for velocity-dependent scattering that has a velocity-dependence as described by Eq.~\ref{eq:veldep}.
In more general terms, our time-step criterion requires that there is a finite velocity for which the fractional velocity change due to the drag force becomes maximal and finite.
This means we are interested in the velocity at which $v \, \sigma_\mathrm{T}(v)$ is maximal.
We remind the reader that the effective drag force for fSIDM was introduced by \cite{Kahlhoefer_2014} and employed to develop a numerical scheme by \cite{Fischer_2021a}.
It is given as
\begin{equation}
    F_\mathrm{drag} = \frac{1}{2} \, \frac{\sigma_\mathrm{T}(v)}{m} \, v^2 \, m_\mathrm{n}^2 \, \Lambda \, .
\end{equation}
The relative particle velocity is denoted by $v$, $m_\mathrm{n}$ is the numerical particle mass and $\Lambda$ is the kernel overlap, a geometrical factor \citep[for details, see][]{Fischer_2021a}.

Assuming the parametrization according to Eq.~\ref{eq:veldep} the fractional velocity change ($\Delta v / v$) due to the drag force becomes maximal for the velocity
\begin{equation} \label{eq:time_step_ve}
    v_e = \frac{w}{\left( -1 -\alpha\right)^{1/\beta}} \, .
\end{equation}
Note that this is only applicable if $\alpha<-1$ and $\beta>0$.
For our choice of $\alpha = -4$ and $\beta = 2$, this implies $v_e = w / \sqrt{3}$.

Using the maximum allowed fractional velocity change $\tau$, we can express the time-step criterion for particle $i$ as

\begin{equation}
    \Delta t_i < \tau \, \frac{2}{v_e} \frac{1}{m_\mathrm{n} \, \Lambda_{ii}} \left(\frac{\sigma_\mathrm{T}(v_e)}{m}\right)^{-1} \, .
\end{equation}
Here, $\Lambda_{ii}$ gives the maximal possible kernel overlap by calculating it with the particle itself.

It is worth pointing out that this time-step depends on the chosen number of neighbours, $N_\mathrm{ngb}$.
With a larger number of neighbours $\Lambda_{ii}$ becomes smaller and thus the time-step is larger and vice versa. Finally, we note that this time-step criterion also applies to rSIDM when using the total cross-section, $\sigma$, instead of $\sigma_\mathrm{T}$.
For rSIDM, the scattering probability reaches a maximum at $v_e$ (see equation~\ref{eq:time_step_ve}) too.

In Appendix~\ref{sec:time_step_disc}, we provide further discussion on issues related to the formulation of a time-step criterion.


\section{Idealized simulations} \label{sec:idealised_sims}

In this section, we present and analyse our idealized simulations and show the results we obtain.
First, we start with a simple test problem in Sec.~\ref{sec:results_thermalisation}.
Secondly, the evolution of the core size for isolated haloes is shown (Sec.~\ref{sec:results_isolated_haloes}) for both initial Hernquist and NFW profiles.

\subsection{Thermalization problem} \label{sec:results_thermalisation}

\begin{table}
    \centering
    \begin{tabular}{l|c|r|c}
        Name & Type & $\sigma_0/m$ & $w$ \\
        & & $[\mathrm{cm}^2 \, \mathrm{g}^{-1}]$ & $[\mathrm{km} \, \mathrm{s}^{-1}]$ \\ \hline
        f10 & Frequent & $10$ & -- \\
        r10 & Rare & $10$ & -- \\
        f4.5e6w0.1 & Frequent & $4.5 \times 10^6$ & $0.1$ \\
        r4.5e6w0.1 & Rare & $4.5 \times 10^6$ & $0.1$
    \end{tabular}
    \caption{
    The table shows the different cross-sections that we used for the thermalization problem. The first column gives the name that we use in the paper to abbreviate the cross-section. It follows the type of self-interaction. Here, ``rare'' corresponds to isotropic scattering. The third column gives $\sigma_0 / m$ and the last one $w$ (see also Eq.~\ref{eq:veldep}).
    }
    \label{tab:cross-sections_therm}
\end{table}

To learn about the differences between a constant and a velocity-dependent cross-section, we first consider the thermalization problem previously studied by \cite{Fischer_2021a}.
This has the advantage that we study the pure effect of DM self-interactions without the influence of gravity.
Hence, it is well suited for the goal of learning about qualitative differences arising from the velocity-dependence.

The numerical set-up consists of a periodic box with a constant density of $10^7 \, \mathrm{M_\odot} \, \mathrm{kpc}^{-3}$ sampled by $10^4$ particles.
The cubic box has a side length of 10 kpc and its particles have a velocity of $2\,\mathrm{km} \, \mathrm{s}^{-1}$ which points into a random direction.
In Tab.~\ref{tab:cross-sections_therm}, we describe the employed cross-sections.
For the velocity-dependent cross-sections, we choose a value for $w$ that is small to have the scattering velocities in the regime where the cross-section decreases strongly with velocity.
The aim is to enhance the qualitative difference between a constant and velocity-dependent cross-section.
While choosing a small value for $w$ we pick a large value for $\sigma_0/m$ to prohibit a drastic increase in the time on which the system evolves compared to the velocity-independent cross-section.

The scattering broadens the velocity distribution such that it evolves towards a Maxwell--Boltzmann distribution.
We can characterize the width of the distribution of the absolute velocities by computing its variance.

\begin{figure}
    \centering
    \includegraphics[width=\columnwidth]{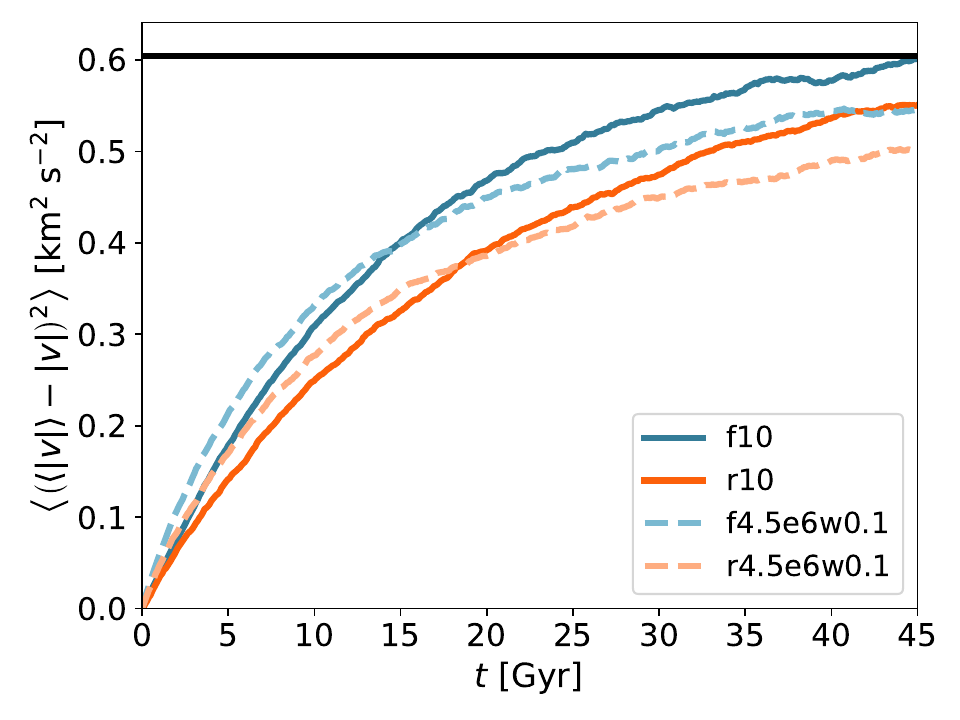}
    \caption{The variance for the distribution of absolute velocities of the thermalization problem introduced by \protect\cite{Fischer_2021a} is shown. We display the results for different SIDM models as a function of time. In black we indicated the variance of the final Maxwell--Boltzmann distribution.
    }
    \label{fig:thermalization}
\end{figure}

In Fig.~\ref{fig:thermalization}, we show the results as a function of time.
For frequent self-interactions, this has been previously studied by \cite{Kasselmann_2021}.
In line with his results, we find that the evolution of the thermalization rate evolves qualitatively differently for velocity-dependent self-interactions compared to a constant cross-section.
The thermalization process evolves faster at early times and slower at late times for the velocity-dependent self-interactions.
For the isotropic cross-section, we find the same.
Initially, the system evolves faster for the velocity-dependent cross-sections, because the cross-section evaluated at the typically relative velocity of the particles is larger compared to the velocity-independent cross-section.
The lower thermalization rate at late times, i.e.\ when the velocity distribution is already close to the Maxwell--Boltzmann distribution, stems mainly from a deviation at the high-velocity tail.
The decrease of the cross-section with velocity makes velocity-dependent self-interactions less efficient in scattering particles to high velocities.
In consequence, the thermalization rate in a late stage is reduced.

\subsection{Isolated haloes} \label{sec:results_isolated_haloes}

Here, we study the evolution of isolated haloes subject to velocity-dependent self-interactions.
Firstly, we investigate the density profile of an isolated halo with a density following a Hernquist profile \citep{Hernquist_1990} and secondly, we do the same for a halo with an NFW profile \citep{Navarro_1996}.
For the two haloes, we also compare rare and frequent self-interactions.

\subsubsection{Hernquist Halo} \label{sec:results_hernquist}

\begin{table}
    \centering
    \begin{tabular}{l|c|r|c}
        Name & Type & $\sigma_0/m$ & $w$ \\
        & & $[\mathrm{cm}^2 \, \mathrm{g}^{-1}]$ & $[\mathrm{km} \, \mathrm{s}^{-1}]$ \\ \hline
        c0 & Collisionless & $0.0$ & -- \\
        f0.8 & Frequent & $0.8$ & -- \\
        r0.8 & Rare & $0.8$ & -- \\
        f1e5w100 & Frequent & $10^5$ & $100$ \\
        r1e5w100 & Rare & $10^5$ & $100$
    \end{tabular}
    \caption{
    The cross-sections that we employed for simulating a Hernquist halo are shown. The columns are the same as in Tab.~\ref{tab:cross-sections_therm}.
    }
    \label{tab:cross-sections_hern}
\end{table}

\begin{figure}
    \centering
    \includegraphics[width=\columnwidth]{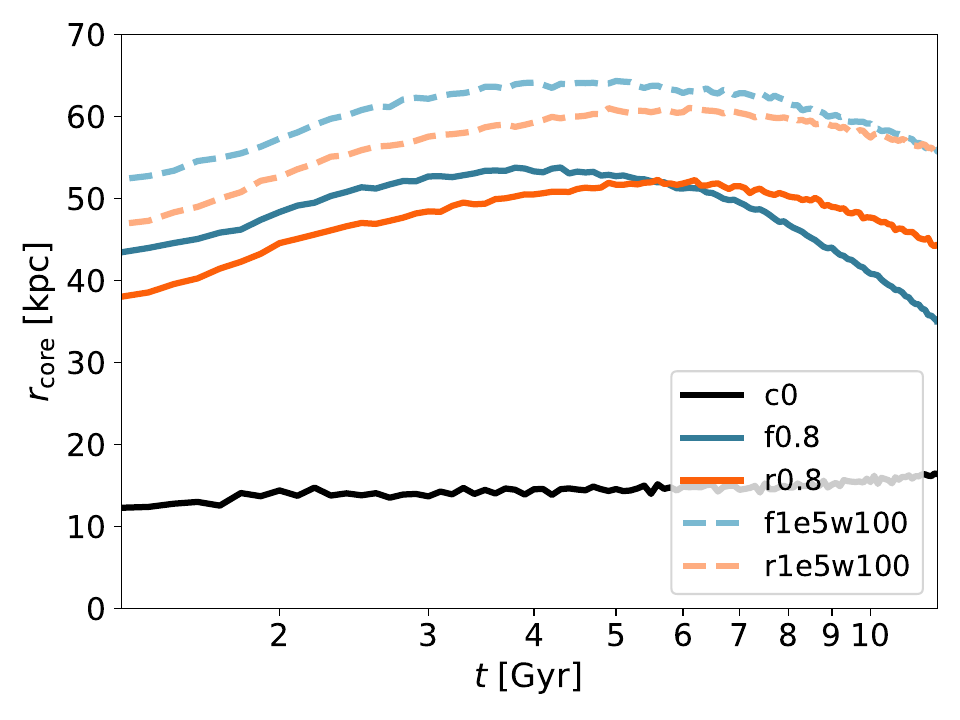}
    \caption{The size of the density core for a Hernquist halo as a function of time is shown when evolved with different DM models. 
    We indicate the cross-section in the legend. The first number refers to $\sigma_0/m$ in units of $\mathrm{cm}^2 \, \mathrm{g}^{-1}$ and the second one to $w$ in units of $\mathrm{km} \, \mathrm{s}^{-1}$ (see Tab.~\ref{tab:cross-sections_hern}). The first two SIDM simulations are for a velocity-independent cross-section and the number gives $\sigma_\mathrm{T} / m$.
    }
    \label{fig:hern_core}
\end{figure}

We simulate the same Hernquist halo as first described by \cite{Robertson_2017b}. It has a mass of $M = 2.46 \times 10^{14} \, \mathrm{M_\odot}$ and a scale radius of $r_s = 279 \, \mathrm{kpc}$.
We generate the initial conditions by sampling the halo up to $r = 400 \, r_s$ using $N=10^7$ particles.
For the gravitational softening length we employ $\epsilon = 0.56 \, \mathrm{kpc}$.
The simulations include velocity-independent and velocity-dependent cross-sections both for fSIDM and rSIDM.
In detail, the cross-sections are shown in Tab.~\ref{tab:cross-sections_hern}.
With this choice, we partially follow \cite{Kasselmann_2021}.
It is worth noting that for the velocity-dependent simulations, the SIDM time-step constraint was tighter than the one from gravity, at least for a fraction of the particles.
This led to a significant increase in computational costs.
We determine the core size, $r_\mathrm{core}$, as previously done by \cite{Robertson_2017a} and \cite{Fischer_2021a} by fitting a cored Hernquist profile.
It is given as
\begin{equation}
    \rho(r) = \frac{M}{2 \uppi} \frac{r_s}{(r^4 + r^4_\mathrm{core})^{1/4}} \frac{1}{(r+r_s)^3}\,.
\end{equation}
As in the original Hernquist profile, $M$ denotes the halo mass and $r_\mathrm{s}$ the scale radius.
To fit the parameters of the density profile we maximize a likelihood based on Poisson statistics,
\begin{equation}
	\mathcal{L} = \prod_i \frac{\lambda_i^{N_i} \, e^{-\lambda_i}}{N_i!} \quad \textnormal{with} \quad \lambda_i = \frac{4\uppi}{m} \int_{r_i}^{r_{i+1}} r^2 \rho(r) \, \mathrm{d}r \,.
\end{equation}
Here, $N_i$ specifies the number of simulation particles in the radial bin $i$, with the boundaries $r_i$ and $r_{i+1}$. This number is compared to the expected value, $\lambda_i$, from the analytic expression of the cored density profile.
For the fit we leave, the core radius, $r_\mathrm{core}$, the scale radius, $r_s$, and the mass, $M$, as free parameters.
Note, this is the same as in \cite{Fischer_2021a}.
The evolution of the core size is shown in Fig.~\ref{fig:hern_core} for the different DM models.\footnote{We found the exact core size to be sensitive to details of the optimization procedure, which might be caused by a noisy likelihood.
This might be the main source of different core sizes for the same halo in the literature \citep{Robertson_2017b, Fischer_2021a, Correa_2022}. Note that \cite{Kochanek_2000} studied the core-size evolution of a Hernquist halo as well, but they employed a different definition of the core size limiting comparability.}

In the early stages, the density core grows due to self-interactions whose effect can be described as heat transfer \citep[e.g.][]{Lynden-Bell_1980, Balberg_2002} that follows the gradient of the velocity dispersion.
As a result, the central region of the halo heats up and its density is decreasing.
For the collisionless DM, we find a small core caused by gravitational two-body interactions, a process known as numerical core formation \citep[e.g.][]{Dehnen_2001}.
At later stages, the core size is decreasing and the halo enters the collapse phase.
In this phase, heat is only transported outward, as the central region cools it also contracts.
Gravitational bound systems are characterized by a negative heat capacity. 
This is for example well known from star clusters but also applies to the haloes we study here.
In consequence, the velocity dispersion at the central region of the halo is increasing and leads to a runaway process called the gravothermal catastrophe.

In previous studies, it was found that the maximum core size that is reached during the haloes evolution is roughly independent of the strength of the cross-section \citep[e.g.][]{Kochanek_2000}, but also its angular dependence \citep[e.g.][]{Robertson_2017a, Fischer_2021a}.
In contrast, we find that the velocity-dependent cross-sections give a larger maximum core size.
However, we have to note that this only occurs for sufficiently small values of $w$.
For the initial Hernquist
halo heat is flowing inwards for radii smaller than the radius of the maximal velocity dispersion, $r({\nu^2_\mathrm{max}})$, this should set the core formation time.
In contrast, for radii larger than $r({\nu^2_\mathrm{max}})$, heat is flowing outwards, determining the core collapse time.
The maximum core size should be a result of the ratio of the total heat in and outflow.
In consequence, a DM candidate that is more efficient in transporting heat inwards than outwards compared to other DM models would produce a larger maximum core size.
We discuss this further in Sec.~\ref{sec:isolated_halo_discussion}, after we have shown the results for the isolated NFW halo.

\begin{figure}
    \centering
    \includegraphics[width=\columnwidth]{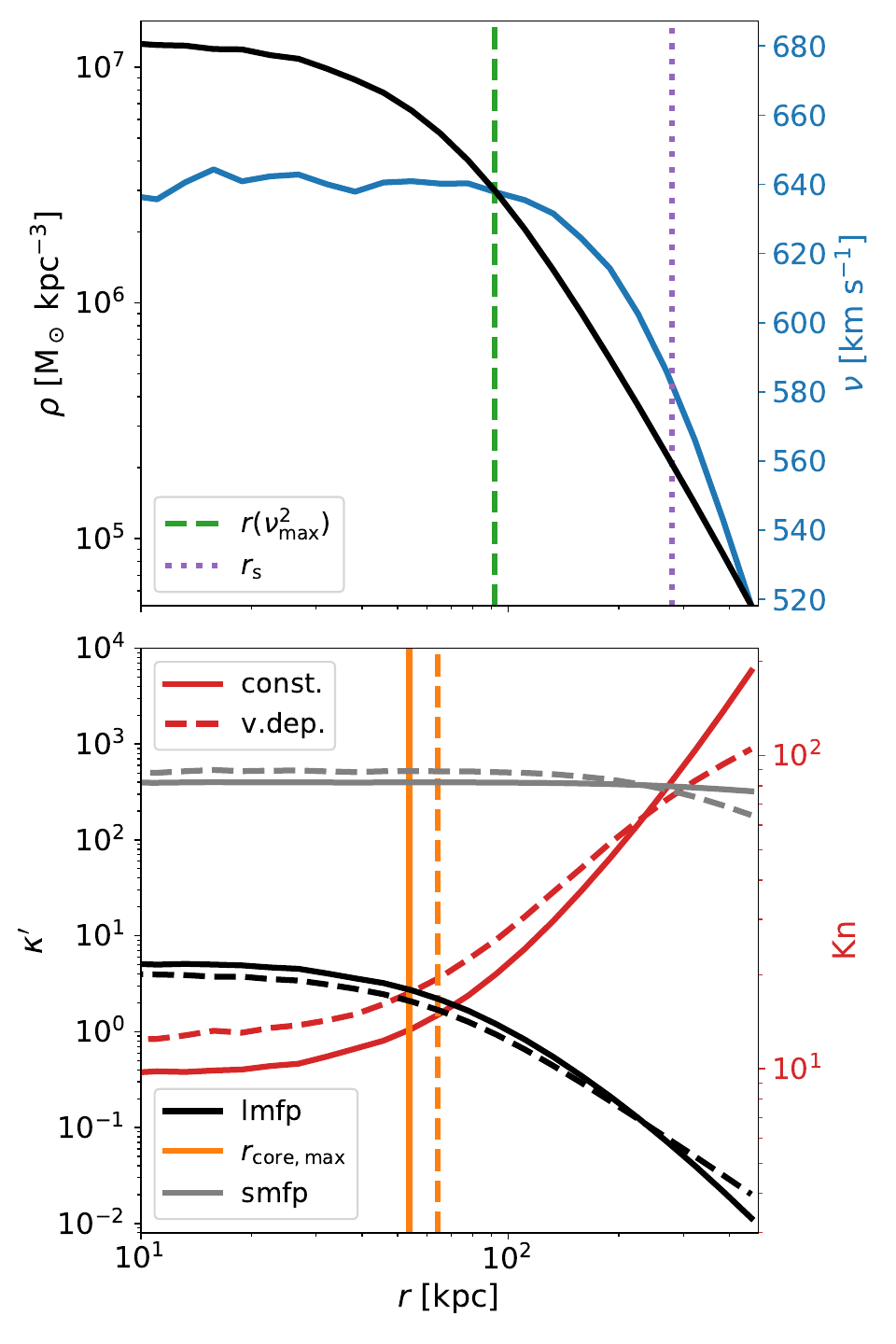}
    \caption{Various properties of the halo following initially a Hernquist profile are shown at the evolution stage when its density core is the largest.
    In the upper panel, we show the density (black) and the velocity dispersion (blue) as a function of radius.
    Moreover, the scale radius, $r_s$, and the radius at which the velocity dispersion of the initial profile reaches its maximum, $r(\nu^2_\mathrm{max})$, are indicated.
    The lower panel gives $\kappa'$ for the smfp (grey) and lmfp (black) regime (see equations~\ref{eq:kappa_smfp} and~\ref{eq:kappa_lmfp}) as well as the Knudsen number (see equation~\ref{eq:knudsen}).
    These quantities are computed based on the effective cross-section, $\sigma_\mathrm{eff}/m$.
    In addition, the maximum core sizes are shown for the runs with the frequent self-interactions, i.e.\ for the velocity-independent and velocity-dependent cross-sections.
    To compute the quantities that are shown as a function of radius, we used the simulation with frequent self-interactions and without velocity dependence.
    }
    \label{fig:hern_det}
\end{figure}

However, to gain further insights into the halo following initially a Hernquist profile, we first plot various quantities at the time of maximum core expansion in Fig.~\ref{fig:hern_det}.
The upper panel shows the density and velocity dispersion profile, and the bottom panel displays quantities related to heat conductivity.

In the following, we describe how we compute the quantities of the bottom panel.
Assuming identical particles the viscosity cross-section is given by
\begin{equation} \label{eq:sigma_v}
    \sigma_v = 4\uppi \int_0^1 \frac{\mathrm{d}\sigma}{\mathrm{d}\Omega} \sin^2 \theta \, \mathrm{d}\cos\theta \,.
\end{equation}
Based on this we can express the effective cross-section of \cite{Yang_2022D} as
\begin{equation} \label{eq:sigma_eff}
    \sigma_\mathrm{eff} = \frac{3}{2} \frac{\langle v^5 \sigma_v(v) \rangle}{\langle v^5 \rangle} \,.
\end{equation}
They introduced the effective cross-section with the aim of matching
differential cross-sections with various angular and velocity dependencies. It thus allows transferring constraints on the strength of self-interactions to various SIDM models.
Here, the average is computed assuming the velocities are well described by a Maxwell–Boltzmann distribution.
Next, we give the heat conductivity using $\sigma_\mathrm{eff}$.
Strictly speaking, we do not specify the heat conductivity $\kappa$, but use $\kappa' = m / \mathrm{k_B} \kappa$, with $m$ the DM particle mass and $\mathrm{k_B}$ the Boltzmann constant.
This is commonly used in the gravothermal fluid model \citep[e.g.][]{Koda_2011}.
Note, \cite{Kummer_2019} took the angular dependence into account by expressing the heat conductivity in terms of the viscosity cross-section.
Here, we go further and use the effective cross-section for $\kappa'$.
For the short-mean-free-path (smfp) regime it is given as
\begin{equation} \label{eq:kappa_smfp}
    \kappa'_\mathrm{smfp} = \frac{9 \, b \, \nu}{4} \left( \frac{\sigma_\mathrm{eff}}{m} \right)^{-1} \quad \textnormal{with} \quad b = \frac{25 \sqrt{\uppi}}{32} \,.
\end{equation}
The one-dimensional velocity dispersion is expressed by $\nu^2$.
In the long-mean-free-path (lmfp) regime, the heat conductivity can be expressed as
\begin{equation} \label{eq:kappa_lmfp}
    \kappa'_\mathrm{lmfp} = \hat{a} \, C \, \frac{\nu^3 \rho}{4 \uppi \mathrm{G}} \left( \frac{\sigma_\mathrm{eff}}{m} \right) \quad \textnormal{with} \quad \hat{a} = \sqrt{\frac{16}{\uppi}}, \, C\approx0.75 \,.
\end{equation}
Here, $\rho$ denotes the density and $\mathrm{G}$ is the gravitational constant.

The Knudsen number, $\mathrm{Kn}$, is usually used to distinguish between the lmfp and smfp regime and is defined as
\begin{equation} \label{eq:knudsen}
    \mathrm{Kn} = \frac{3}{2} \, \sqrt{\frac{4\uppi \mathrm{G}}{\rho \nu^2}} \left( \frac{\sigma_\mathrm{eff}}{m} \right)^{-1} \;.
\end{equation}
Numerically $\mathrm{Kn}>1$ corresponds to the lmfp regime and $\mathrm{Kn}<1$ to the smfp regime.

From the upper panel of Fig.~\ref{fig:hern_det}, we can see that the velocity dispersion at the time of maximum core expansion is roughly constant for radii smaller than $r(\nu^2_\mathrm{max})$. However, the density core itself is much smaller, resulting in a steep density gradient at $r(\nu^2_\mathrm{max})$.

The bottom panel shows the maximum core sizes for the velocity-dependent and velocity-independent cross-sections. It is visible that the maximum core size is smaller than $r(v_\mathrm{max}^2)$.
Moreover, we can see that the Knudsen number is increasing as a function of radius and is always much larger than unity. Implying that the halo is always in the lmfp regime.
For radii smaller than $r_\mathrm{s}$, the corresponding heat conductivity ($\kappa'_\mathrm{lmfp}$) is larger for the velocity-independent cross-section. In contrast, $\kappa'_\mathrm{smfp}$ is larger for the velocity-dependent cross-section.
If the cross-section is decreasing as a function of velocity, smaller scattering velocities may play a more important role compared to large velocities in the heat conduction than for velocity-independent cross-sections (see also Sec.~\ref{sec:results_thermalisation}).

However, using the effective cross-section may eventually be problematic for extreme velocity dependencies.
Depending on the velocity of a DM particle, it sees different distributions of relative velocities and thus has a mean free path that depends on its velocity.
Unfortunately, it is not understood how the evolution in the lmfp regime could be derived from first principles.
This complicates a precise description of the heat conduction in the halo.

\subsubsection{NFW Halo} \label{sec:results_nfw}

\begin{table}
    \centering
    \begin{tabular}{l|c|r|c}
        Name & Type & $\sigma_0/m$ & $w$ \\
        & & $[\mathrm{cm}^2 \, \mathrm{g}^{-1}]$ & $[\mathrm{km} \, \mathrm{s}^{-1}]$ \\ \hline
        c0 & Collisionless & $0$ & -- \\
        f10 & Frequent & $10$ & -- \\
        r10 & Rare & $10$ & -- \\
        f5e3w720 & Frequent & $5 \times 10^3$ & $720$ \\
        r5e3w720 & Rare & $5 \times 10^3$ & $720$ \\
        f2.5e5w180 & Frequent & $2.5 \times 10^5$ & $180$ \\
        r2.5e5w180 & Rare & $2.5 \times 10^5$ & $180$
    \end{tabular}
    \caption{The cross-sections that we employed for simulating an NFW halo are shown. The columns are the same as in Tab.~\ref{tab:cross-sections_therm}.
    }
    \label{tab:cross-sections_nfw}
\end{table}

We studied the core formation in an isolated NFW halo using various DM models.
These include velocity-independent cross-sections for fSIDM and rSIDM each with $\sigma_\mathrm{T} / m = 10.0 \, \mathrm{cm}^2 \mathrm{g}^{-1}$ and velocity-dependent fSIDM and rSIDM cross-section with $\sigma / m = 5000.0 \, \mathrm{cm}^2 \mathrm{g}^{-1}$, $w = 720 \, \mathrm{km} \, \mathrm{s}^{-1}$ and $\sigma / m = 2.5 \times 10^5 \, \mathrm{cm}^2 \mathrm{g}^{-1}$, $w = 180 \, \mathrm{km} \, \mathrm{s}^{-1}$.
The cross-sections and the abbreviations we use for them are also shown in Tab.~\ref{tab:cross-sections_nfw}.

For the NFW halo, we use the same initial conditions as used by \cite{Fischer_2021a} for their fig.~5.
Our halo has a virial mass of $10^{15}\,\mathrm{M_\odot}$, a scale radius of $300 \, \mathrm{kpc}$ and a density parameter of $\rho_\mathrm{0} \equiv 4 \rho(r_\mathrm{s}) = 2.9 \times 10^6 \, \mathrm{M_\odot \, kpc}^{-3}$.
The halo is sampled up to the virial radius ($r_\mathrm{vir} = 1626 \, \mathrm{kpc}$) and resolved by $N=10^6$ particles.
For the simulations we employ a gravitational softening length of $\epsilon = 0.56\, \mathrm{kpc}$.

\begin{figure}
    \centering
    \includegraphics[width=\columnwidth]{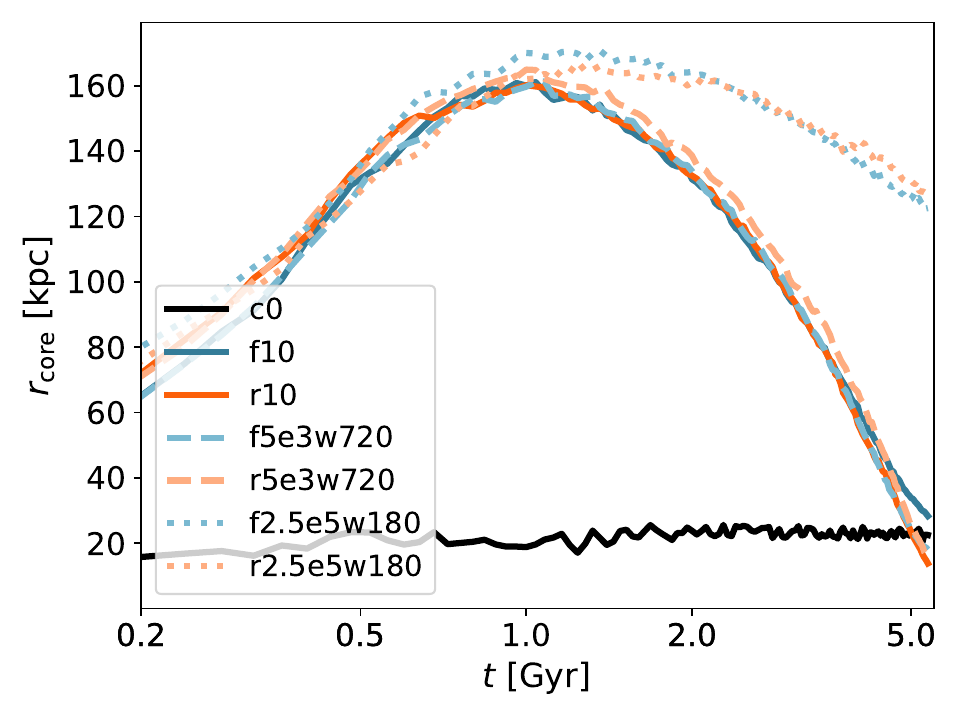}
    \caption{We display the core size for an NFW halo, which we simulated with different DM models.
    The abbreviations for the cross-sections are explained in Tab.~\ref{tab:cross-sections_nfw}.
    }
    \label{fig:nfw_core}
\end{figure}

\begin{figure}
    \centering
    \includegraphics[width=\columnwidth]{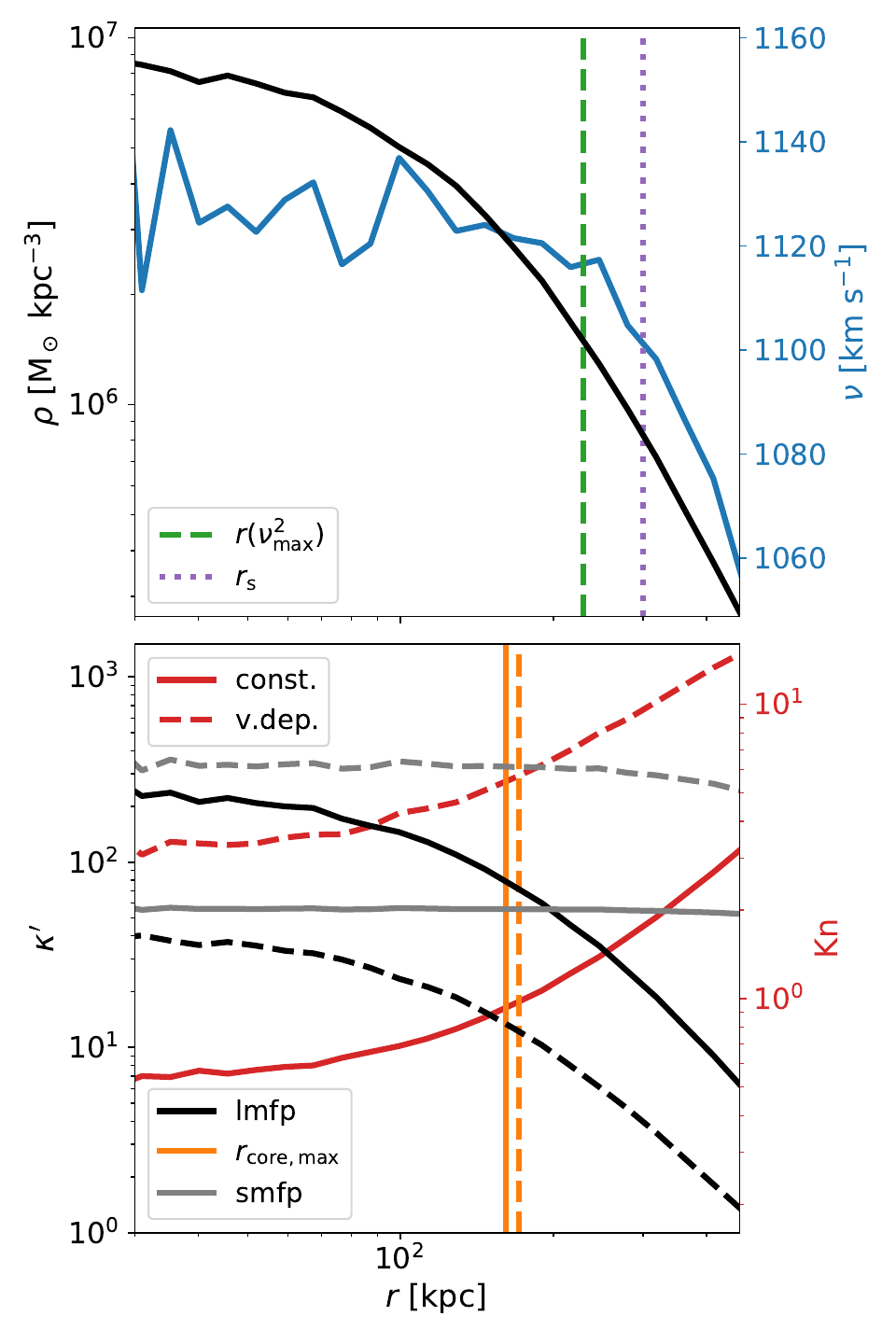}
    \caption{We show the same as in Fig.~\ref{fig:hern_det}, but for the NFW simulations. The maximum core size of the velocity-dependent model refers to the run with f2.5e5w180.
    }
    \label{fig:nfw_det}
\end{figure}

We measure the core size by fitting a cored NFW profile.\footnote{In the literature also other descriptions of a cored NFW profile exist \citep[e.g.][]{Read_2016a, Read_2016b, Ray_2022}. The one we use corresponds to the one employed by \cite{Yang_2023D} for their parametric model of the evolution of a halo following initially an NFW profile.} It is given by
\begin{equation} \label{eq:likelihood}
    \rho(r) = \frac{\rho_0}{( r^4 + r^4_\mathrm{core})^{1/4}} \, \frac{r_\mathrm{s}}{\left( 1+r/r_\mathrm{s} \right)^2}\, .
\end{equation}
For the fitting procedure, we have $\rho_0$, $r_\mathrm{s}$, and $r_\mathrm{core}$ as free parameters. We maximize a likelihood based on Poisson statistics (equation~\ref{eq:likelihood}) as described in section~4 of \cite{Fischer_2021a}.

The core sizes for different DM models are shown in Fig.~\ref{fig:nfw_core}.
First, we consider the cross-sections, f10, r10, f5e3w720, and r5e3w720.
For the phase of the core formation and the onset of core collapse up to $\approx 4 \, \mathrm{Gyr}$, the core sizes are very similar.
Only the velocity-dependent rSIDM cross-section yields slightly larger core sizes.
Hence, the momentum transfer cross-section provides a good match between fSIDM and rSIDM in the given case.
Only at later stages of the halo evolution do differences between the models occur.
When the core size is almost zero, it seems that small-angle scattering slows down the core collapse compared to isotropic scattering.
These results are partially in line with previous work.
\cite{Yang_2022D} found that a constant and velocity-dependent cross-section behave qualitatively very similarly for most of the halo evolution but differ at the late stages of the collapse phase.
They also found that the viscosity cross-section provides a better match between different angular dependencies than the momentum transfer cross-section.
In the companion paper \citep[][]{Sabarish_2023}, it is found that the viscosity cross-section can indeed provide a reasonable, but not perfect match between isotropic scattering and a very anisotropic cross-section in the fSIDM limit.
In contrast, for our set-up with a much stronger cross-section the momentum transfer cross-section provides a very good match  regardless of the velocity dependence.
However, we should point out that the quality of the match depends on the halo properties and the strength of the self-interactions \citep[see e.g.\ fig.~9 of][we show this result again in Sec.~\ref{sec:results_density_profile}]{Fischer_2022}.
Here, one can see for the larger cross-section that the momentum transfer cross-section match yields a larger effect of fSIDM on the central densities of DM haloes at the high-mass end compared to rSIDM.
For lower-mass haloes, it changes and rSIDM has a stronger effect on the central halo density.
As \cite{Yang_2022D} simulated NFW haloes with a mass of $M_\mathrm{200} \approx 10^7 \, \mathrm{M_\odot}$ and a concentration of $c_\mathrm{200} \approx 20$ (for details see their table~1), they probed a different regime than we do here.
Hence, the quality of a matching procedure for the angular dependence could depend on the halo properties and the strength of the self-interactions.
It is also important to note that the inner regions of our NFW halo are in the smfp regime or close to it ($\mathrm{Kn}<1$) and not in the lmfp regime for the velocity-independent cross-sections.

For the strongly velocity-dependent cross-section, i.e.\ the one with $w = 180\,\mathrm{km} \, \mathrm{s}^{-1}$, we find that the evolution differs qualitatively from the ones with a weaker velocity dependence.
The results are somewhat similar to the results for the Hernquist halo, the maximum core size becomes larger and the collapse time longer compared to the core formation time.
However, the increase in the maximum core size is weaker compared to the Hernquist halo.
This could be because the cross-section we have simulated is not as extremely velocity-dependent as for the Hernquist halo ($w=100 \, \mathrm{km} \, \mathrm{s}^{-1}$ for the Hernquist halo and $w=180 \, \mathrm{km} \, \mathrm{s}^{-1}$ for the NFW halo).
Note that the NFW halo has a larger total mass and hence a larger velocity dispersion than the Hernquist halo, such that the two simulations cannot be directly compared.
But when $w$ is compared to the typical scattering velocity of the halo, the velocity dependence appears to be similar.
In consequence, it is plausible that the difference in maximum core size stems primarily from a different reason such as the details of the density profile.

Analogous to the Hernquist halo we have computed the same quantities as in Fig.~\ref{fig:hern_det}, but for the NFW halo and show them in Fig.~\ref{fig:nfw_det}.
In contrast to the Hernqusit halo, we find that the central region of the halo has a Knudsen number smaller than unity when simulated with the velocity-independent cross-section and thus would be considered to be in the smfp regime.
In addition, the heat conductivity in the two regimes is more similar.
But the Knudsen number varies strongly with velocity dependence.
As for the Hernquist halo $\kappa'_\mathrm{smfp}$ has a larger value in the case of the velocity-dependent cross-section and $\kappa'_\mathrm{lmfp}$ is larger for the velocity-independent cross-section.

\subsubsection{Discussion of isolated halo evolution} \label{sec:isolated_halo_discussion}

In this last part on isolated haloes, we discuss the physics driving their evolution. 
During the evolution of the halo, the central velocity dispersion is increasing and the effective strength of the self-interactions may change according to the velocity dependence of the cross-section. 
An increasing velocity dispersion implies higher relative velocities of the DM particles and for a cross-section that decreases with velocity this leads to fewer scatterings.

The halo may reach its maximum core size when the gradient of the velocity dispersion has become zero.
Afterwards, heat is only flowing outwards, which leads to a shrinking density core and the gravothermal collapse of the halo. While the density core is shrinking the central velocity dispersion is increasing.
Given this increase in velocity dispersion, one would expect that the collapse is slowing down for a velocity-dependent cross-section compared to a velocity-independent one.
However, in our simulations, we do not find an indication that the rate at which the density core is shrinking changes due to the velocity dependence (see Fig.~\ref{fig:nfw_core}).
Instead, we only found that the core collapse time scale relative to the core formation time scale changes.

The evolution of the halo may not only be determined by the central region but also by larger radii, at least radii up to $r(\nu^2_\mathrm{max})$ and a bit beyond may play a crucial role.
A core-collapse rate that is insensitive to the velocity dependence might be caused by the relevant velocity dispersion staying roughly constant.
Indeed the velocity dispersion at larger radii is less affected during the evolution and may play a crucial role in the core collapse. 
Right from the beginning of the simulation, during core formation, heat flows outward at radii larger than $r(\nu^2_\mathrm{max})$.
This heat flow takes place at velocities that are larger than in the central region of the halo.
In consequence, the ratio of heat inflow and outflow depends on the velocity dependence of the scattering.
For example, this is visible in the core formation and core collapse times. They are set by heat inflow and outflow.

The cross-sections we have simulated lead to roughly the same core formation time.
For strongly velocity-dependent cross-sections, less heat outflow takes place during that time.
This can result in a larger maximum core size as we found for the Hernquist halo (see Fig.~\ref{fig:hern_core}).
The maximum core size depends on the transition radius between heat inflow and outflow.
Initially, this radius is set by $r(\nu^2_\mathrm{max})$ but evolves according to the ratio of heat in and outflow.
As we found, this evolution is only significantly affected by strongly velocity-dependent cross-sections.

Overall it becomes clear that if the scattering is velocity dependent, the evolution of an isolated halo can change qualitatively.
However, we do not have a precise understanding of the physical mechanisms driving this difference.
How effective the heat outflow taking place in the lmfp regime could depend on the gradient of the gravitational potential and the ability to scatter particles to large velocities.
It could be mainly the high-velocity particles exceeding the escape velocity and carrying energy away that drive the core collapse.
In this context, the exact density profile may eventually matter.
For example the Hernquist and NFW profiles that we have investigated, have a different slope in the outskirts.
Implying a different gradient of the gravitational potential.
Further investigation is needed to fully understand the evolution of isolated haloes.


\section{Cosmological Simulations} \label{sec:cosmo_sims}

We present our cosmological simulations in this section and show the results we obtain.
First, we describe the simulations, followed by the analysis of the data.
This includes many aspects such as the density and shape profiles of the DM haloes and the abundance of satellites.

\subsection{Simulations} \label{sec:cosmo_sims_sims}

\begin{figure}
    \centering
    \includegraphics[width=\columnwidth]{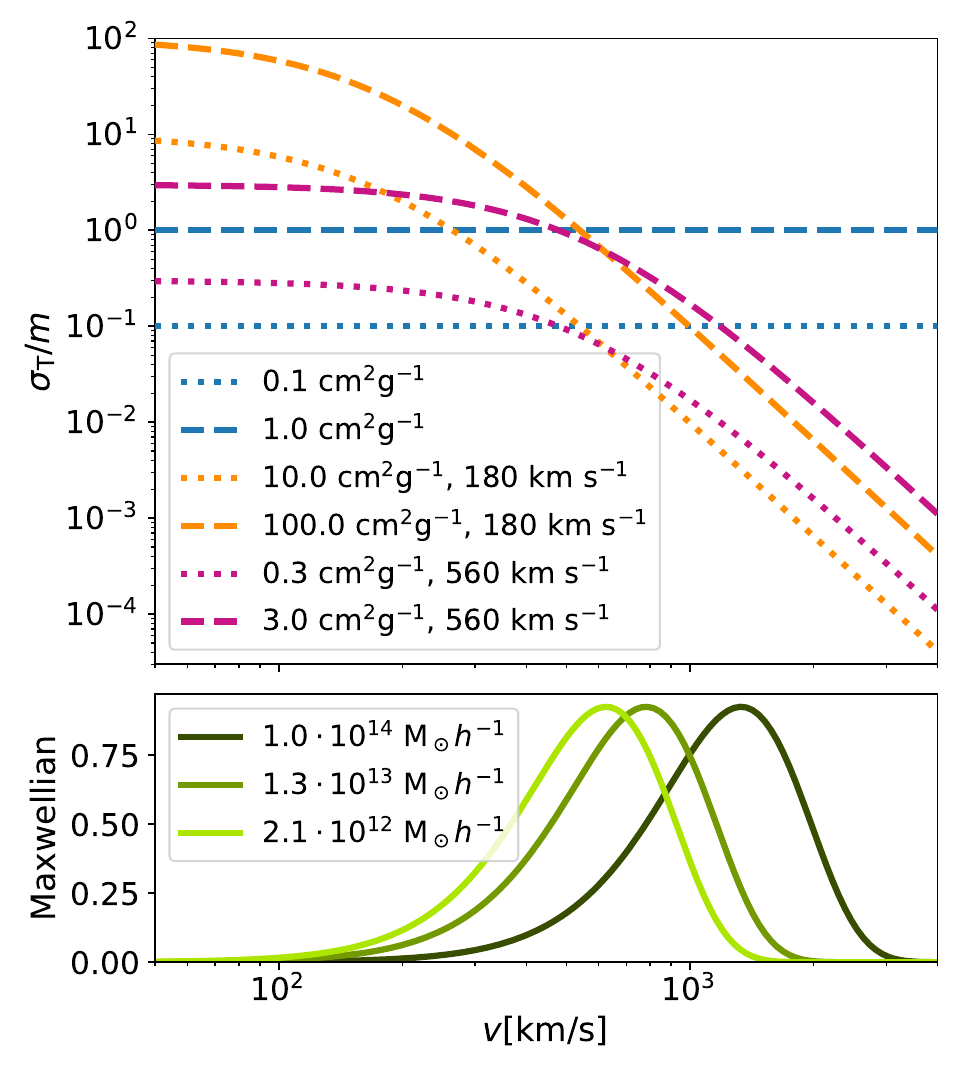}
    \caption{In the upper panel, we illustrate the cross-sections used for our cosmological simulations.
    In blue, we show the velocity-independent cross-sections from \citet{Fischer_2022}.
    The velocity-dependent cross-sections are displayed in orange ($w = 180 \,\mathrm{km}\,\mathrm{s}^{-1}$) and purple ($w = 560 \,\mathrm{km}\,\mathrm{s}^{-1}$).
    In the lower panel, we indicate in green typical scattering velocities.
    The Maxwell-Boltzmann distributions (see equation~\ref{eq:maxwellian}) correspond to the scattering velocities in the centres of the haloes from the three halo mass bins that we use in Sec.~\ref{sec:cosmo_sims_results}.
    }
    \label{fig:cross-sections}
\end{figure}

We have run several simulations of CDM, rSIDM, and fSIDM.
For the SIDM models, we use two different velocity dependencies, namely $w=180\, \mathrm{km} \, \mathrm{s}^{-1}$ and $w=560 \,\mathrm{km} \, \mathrm{s}^{-1}$.
For each of them we have models that differ in $\sigma_0$ by one order of magnitude.
Our simulations are run with fSIDM and a momentum transfer matched isotropic cross-section.
The details of the DM models are given in Tab.~\ref{tab:cross-sections_cosmo} and their velocity-dependence is plotted in Fig.~\ref{fig:cross-sections}.
Here, we also show the scattering velocities inside the centres of haloes from three different mass bins, which we use in Section~\ref{sec:cosmo_sims_results}. The velocities are indicated with a Maxwell--Boltzmann distribution that runs logarithmically in velocity:
\begin{equation} \label{eq:maxwellian}
    f_\mathrm{log}(v_\mathrm{scat}) = \sqrt{\frac{2}{\uppi}} \frac{v_\mathrm{scat}^3}{a^3} \, e^{-\frac{v_\mathrm{scat}^2}{2 \, a^2}} \quad \textnormal{with} \quad a = \sqrt{2 \, \nu^2} \,.
\end{equation}
The distribution of scattering velocities, $v_\mathrm{scat}$, depends on the one-dimensional velocity dispersion, $\nu^2$, of the halo.
In Appendix~\ref{sec:constraints}, we put those DM models in the context of current observational constraints on the strength of DM self-interactions.

\begin{table}
    \centering
    \begin{tabular}{l|c|r|c}
        Name & Type & $\sigma_0/m$ & $w$ \\
        & & $[\mathrm{cm}^2 \, \mathrm{g}^{-1}]$ & $[\mathrm{km} \, \mathrm{s}^{-1}]$ \\ \hline
        c0 & Collisonless & $0.0$ & -- \\
        f0.1 & Frequent & $0.1$ & -- \\
        r0.1 & Rare & $0.1$ & -- \\
        f1 & Frequent & $1.0$ & -- \\
        r1 & Rare & $1.0$ & -- \\
        f10w180 & Frequent & $10.0$ & $180$ \\
        r10w180 & Rare & $10.0$ & $180$ \\
        f100w180 & Frequent & $100.0$ & $180$ \\
        r100w180 & Rare & $100.0$ & $180$ \\
        f0.3w560 & Frequent & $0.3$ & $560$ \\
        r0.3w560 & Rare & $0.3$ & $560$ \\
        f3w560 & Frequent & $3.0$ & $560$ \\
        r3w560 & Rare & $3.0$ & $560$
    \end{tabular}
    \caption{The table shows the different cross-sections that we used for the cosmological simulations.
    Analogously to Tab.~\ref{tab:cross-sections_therm}, we use the same columns.
    Note, that the simulations of the first five DM models have been presented by \citet{Fischer_2022}.
    }
    \label{tab:cross-sections_cosmo}
\end{table}

For the full box cosmological simulations, we use the same ICs as by \cite{Fischer_2022}.
They are similar to box 4 of the Magneticum simulations\footnote{Magneticum: \url{http://www.magneticum.org}} and have a comoving side length of $48 \, \mathrm{Mpc} h^{-1}$.
The employed cosmological model is described by the following parameters: $\Omega_\mathrm{M} = 0.272$, $\Omega_\mathrm{\Lambda} = 0.728$, $h=0.704$, $n_s = 0.963$, and $\sigma_8 = 0.809$ \citep[WMAP7;][]{Komatsu_2011}.
Further properties can be found in Tab.~\ref{tab:cosmo_sims}.

\begin{table}
    \centering
    \begin{tabular}{c|c|c|c|c}
        Name & $l_\mathrm{box}$ & $N_\mathrm{DM}$ & $m_\mathrm{DM}$ \\
        & $(\mathrm{cMpc} \, h^{-1})$ &    & $(\mathrm{M_\odot} \, h^{-1})$ \\ \hline
        hr & 48 & $216^3$ & $8.28 \times 10^8$\\
        uhr & 48 & $576^3$ & $4.37 \times 10^7$ 
    \end{tabular}
    \caption{The table gives the different simulations we run. The first column denotes the name, the second one the box size, the third one the number of numerical DM particles and the last one the mass of the numerical DM particles.
    Each set-up was run with eight different velocity-dependent cross-sections as described in Tab.~\ref{tab:cross-sections_cosmo}.}
    \label{tab:cosmo_sims}
\end{table}

The DM haloes are identified using the friends-of-friends algorithm,\footnote{A description of the friends-of-friends algorithm can, for example, be found in the work by \cite{More_2011}.} which is implemented in \textsc{opengadget3}.
The mass of a halo, $M$, is computed as the sum of the gravitationally bound particles.
The virial radius, $r_\mathrm{vir}$, and the virial mass, $M_\mathrm{vir}$, are measured with the spherical-overdensity approach based on the overdensity predicted by the generalized spherical top-hat collapse model \cite[e.g.][]{Eke_1996}.
Here, $r_\mathrm{vir}$ is defined as the radius at which the mean density becomes larger than the one of the top-hat collapse model and $M_\mathrm{vir}$ is the mass inside $r_\mathrm{vir}$.

We use \textsc{SubFind} \citep{Springel_2001, Dolag_2009}, which is implemented as part of \textsc{opengadget3}, to identify the substructure in the simulation.
Every halo contains at least one subhalo, which is the primary subhalo located at the same position as the halo (determined by the location of the most gravitationally bound particle of the halo).
The primary subhalo typically contains most of the particles that belong to the halo, but this is not necessarily the case.

\subsection{Results} \label{sec:cosmo_sims_results}
In the following, we show the results of our cosmological simulations.
The simulation set-up we used is described in Sec.~\ref{sec:cosmo_sims_sims}.
We begin with the surface density of a massive halo (Section~\ref{sec:results_surface_density}).
Subsequently, we discuss the density profiles of the haloes in Section~\ref{sec:results_density_profile} and continue with their shapes (Section~\ref{sec:results_shapes}).
We investigate the abundance of satellites (Section~\ref{sec:results_satellites}) as well as their diversity in terms of the circular velocity (Sec.~\ref{sec:results_diversity}).
Finally, in Sec.~\ref{sec:results_frequent_vs_rare}, we study differences between frequent and rare self-interactions in the context of velocity-dependent scattering.

\begin{figure*}
    \centering
    \includegraphics[width=\textwidth]{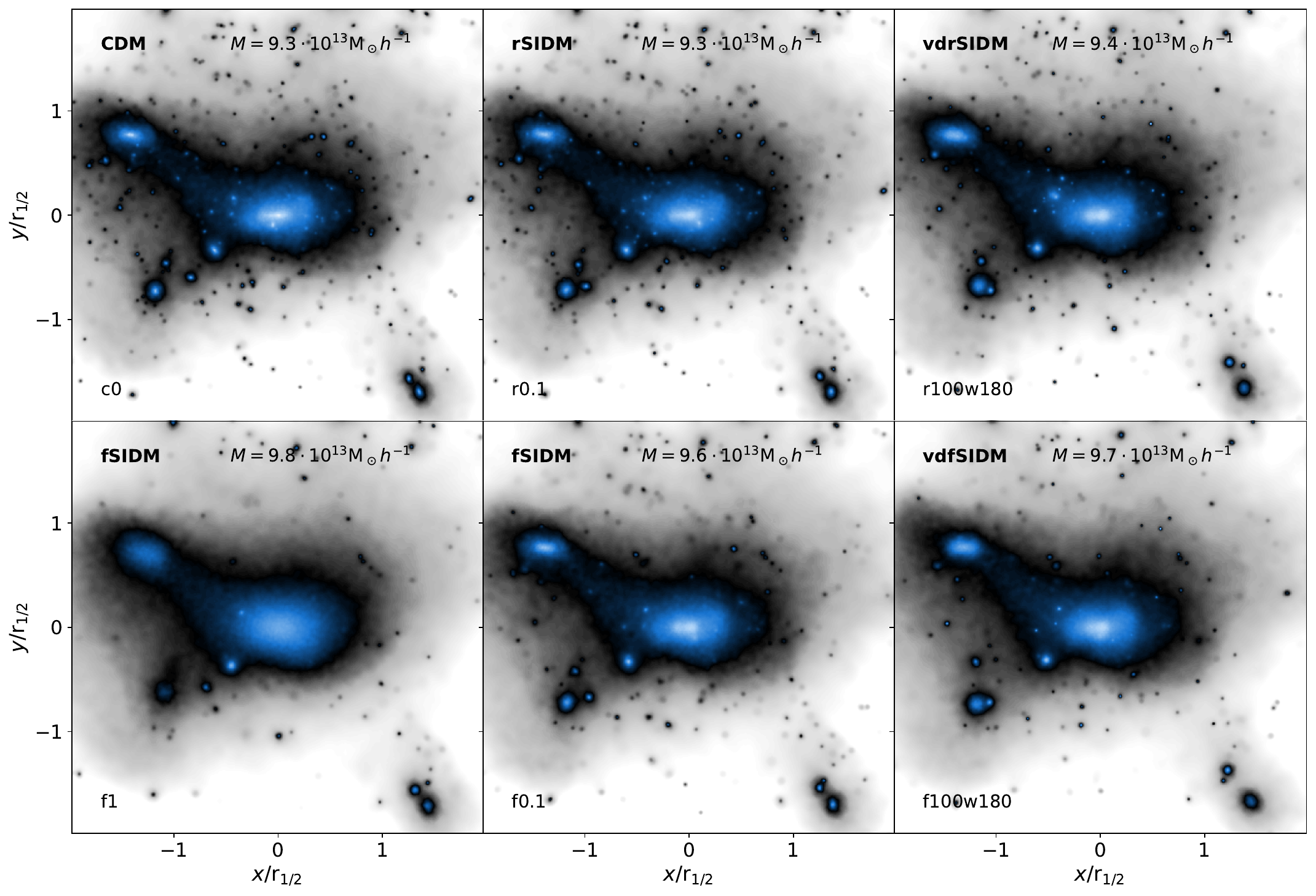}
    \caption{The surface density of the fourth most massive system in our simulation is shown. We cross-identified it among all simulations and show it from the same perspective. We rotate the system such that for CDM the semimajor axis is parallel to the $x$-axis and the semiminor axis parallel to the $y$-axis. We scale the axes in terms of $r_{1/2}$, the half mass radius of the primary subhalo in the CDM simulation. The surface density is indicated with a logarithmic colour scaling. We use the same for each panel. The abbreviation of the cross-section is given in the lower left corner of each panel and the detailed parameters can be looked up in Tab.~\ref{tab:cross-sections_cosmo}.}
    \label{fig:map}
\end{figure*}

\subsubsection{Surface Density} \label{sec:results_surface_density}

In Fig.~\ref{fig:map}, we show the surface density of the same halo but in different DM models.
It is the fourth most massive halo ($M = 9.3 \times 10^{13} \, \mathrm{M_\odot} h^{-1}$) in our simulation and nicely illustrates the effects of SIDM.
They are most pronounced when comparing the two panels on the left-hand side, as the fSIDM simulation of the two has relatively strong self-interactions ($\sigma_\mathrm{T} / m = 1.0 \, \mathrm{cm}^2 \, \mathrm{g}^{-1}$).
Typical effects of SIDM that can be seen here are the formation of a density core, the rounder shape of haloes and the suppression of substructure.
Many of the satellites visible in the CDM run do not exist in the fSIDM run.
However, in the other SIDM runs shown here the suppression of the satellite abundance is weaker.
There exist even objects for which no counterpart in the CDM simulation can be identified by eye.
This is in particular the case for the velocity-dependent cross-section shown in the right-hand side panels.
In the following sections, we quantify these self-interaction-induced changes in the DM distribution.

\subsubsection{Density Profiles} \label{sec:results_density_profile}

\begin{figure*}
    \centering
    \includegraphics[width=\textwidth]{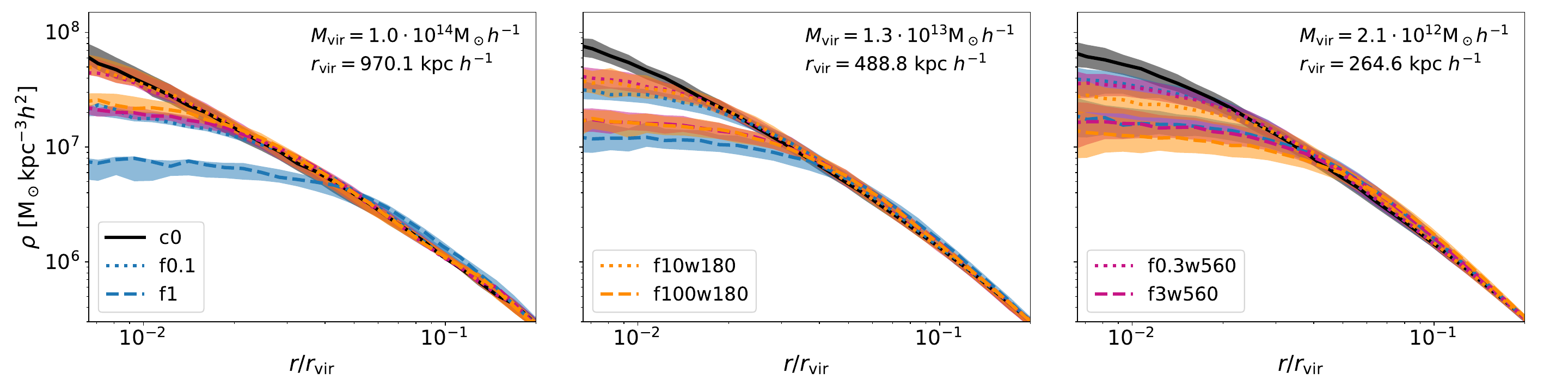}
    \caption{We show the median density profile for haloes from three different mass bins.
    The results for the velocity-independent and velocity-dependent cross-sections are displayed together.
    However, we show the results only for fSIDM as the rSIDM results are similar.
    The density is plotted as a function of the radius in units of the virial radius.
    The shaded regions indicate the scatter among the haloes, and the range between the 25th and 75th percentiles is displayed.
    The virial mass and the virial radius given in the panels indicate the median of the corresponding mass bin from the CDM simulation.
    All plots show the profiles for a redshift of $z = 0$ and are produced from the full cosmological box with the highest resolution.
    Note, we have used all particles, not only those that belong to the halo as identified by \textsc{SubFind}.}
    \label{fig:density_profile}
\end{figure*}

\begin{figure*}
    \centering
    \includegraphics[width=\textwidth]{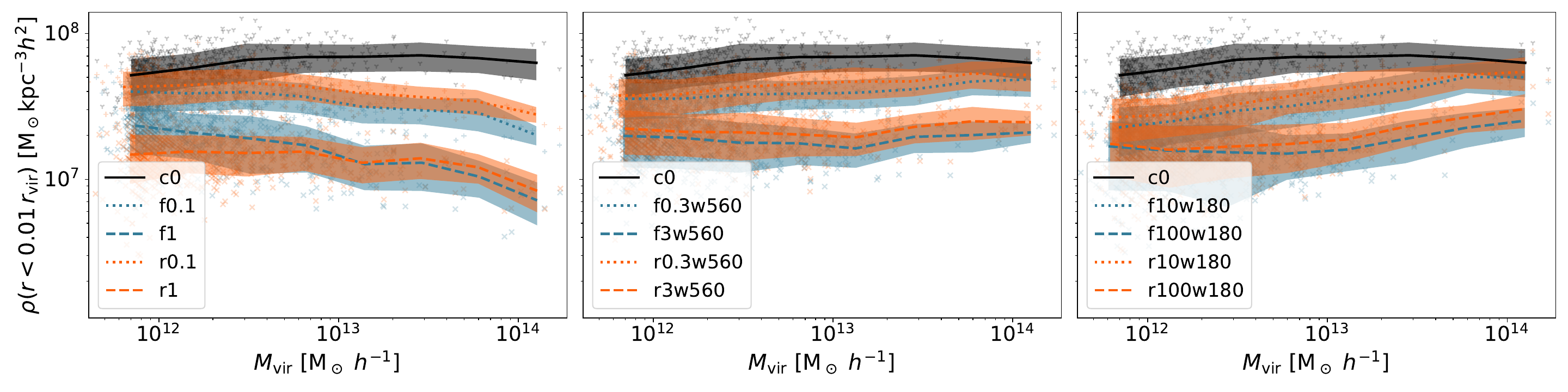}
    \caption{The central density of the DM haloes is shown as a function of their virial mass.
    We measure the central density as the mean density within a radius of $0.01 r_\mathrm{vir}$.
    In the left-hand panel, the simulations with a velocity-independent cross-section are shown \citep[reprint of fig.~9 of][]{Fischer_2022}.
    The middle panel gives the velocity-dependent scattering with $w=560 \, \mathrm{km}\,\mathrm{s}^{-1}$ and the right-hand panel displays the self-interactions with $w=180 \, \mathrm{km}\,\mathrm{s}^{-1}$.
    Individual systems are indicated by ``+’’ when evolved with the smaller cross-section.
    For the larger cross-section, we use ``$\times$'' and the CDM case is marked by ``$\downY$''. In addition, we computed the mean of the distribution as a function of virial mass, shown by the lines. The shaded regions give the corresponding standard deviation.
    }
    \label{fig:central_density}
\end{figure*}

A quantity commonly measured for SIDM is the density profile of haloes.
In particular, the formation of a central density core that is characterized by a shallow gradient and a lower density compared to CDM \citep[except][]{ONeil_2023}. We have studied this in an idealized set-up in Section~\ref{sec:idealised_sims}.
Within the cosmological context this has been measured by various authors \citep[e.g.][]{Stafford_2020, Eckert_2022, Mastromarino_2023} and used to constrain the strength of DM self-interactions (see Appendix~\ref{sec:constraints}).

We investigate the DM density profile for the haloes of our cosmological simulations.
In particular, we study the median density profile within three halo mass bins.
This is shown in Fig.~\ref{fig:density_profile}, where we indicated the median virial mass and virial radius of the haloes contained in the three mass bins.
We show all cross-sections we have simulated the ones with $w=180 \, \mathrm{km} \, \mathrm{s}^{-1}$ are shown in orange and the ones with $w=560 \, \mathrm{km} \, \mathrm{s}^{-1}$ are shown in purple.
The small cross-sections, i.e. the one with the smaller $\sigma_0/m$ for each $w$ show hardly any core formation for the most massive haloes (left-hand panel).
But for the less massive haloes, the core size is increasing in terms of the virial radius, $r_\mathrm{vir}$.
This is a consequence of the relative velocities between the DM particles being smaller for less massive systems.
As a result, the particles typically scatter at smaller relative velocities for which the interaction strength is larger compared to high velocities (see also Fig.~\ref{fig:cross-sections}).

While two cross-sections with a different velocity dependence can behave similarly at a specific mass scale they may vastly differ at another mass scale.
However, their qualitative behaviour is similar for relaxed systems, i.e.\ in our model it would be possible to find a different value for $\sigma_0/m$ that resembles the behaviour of a cross-section with a vastly different value for $w$.
This allows transferring constraints between models of different velocity dependencies and gave rise to the effective cross-section (see equation~\ref{eq:sigma_eff}) introduced by \cite{Yang_2022D}.

In Fig.~\ref{fig:central_density}, we show the central density of the DM haloes as a function of their virial mass.
For the velocity-independent cross-section (left-hand panel), we find that it is decreasing as a function of halo mass when self-interactions are present.
When considering the velocity-dependent runs it becomes clear that the gradient with halo mass depends on the velocity-dependence of the self-interactions.
For $w = 560 \, \mathrm{km} \, \mathrm{s}^{-1}$ there is no or only a weak trend with halo mass (middle panel).
But for the $w = 180 \, \mathrm{km} \, \mathrm{s}^{-1}$ cross-section (right-hand panel), the central density is increasing with halo mass and thus the trend is opposite to the simulations with a constant cross-section.

Note that we used the momentum transfer cross-section to match rSIDM and fSIDM.
If we would have used the viscosity cross-section, the fSIDM cross-section would only have 2/3 of its value to correspond to the simulated rSIDM cross-section.
A detailed derivation of this factor has been presented by \citep{Sabarish_2023}.
This would imply larger central densities for the fSIDM cross-sections.
In consequence, it probably would often provide a better matching.
Except for haloes with masses lower than $M_\mathrm{vir} \approx 10^{13} \mathrm{M_\odot}$ and simulated with the strong and velocity-independent scattering.
Here, the matching would become worse.
It should be noted, that not all haloes used in Fig.~\ref{fig:central_density} are relaxed which makes the picture more complicated.

\subsubsection{Shapes} \label{sec:results_shapes}

\begin{figure*}
    \centering
    \includegraphics[width=\textwidth]{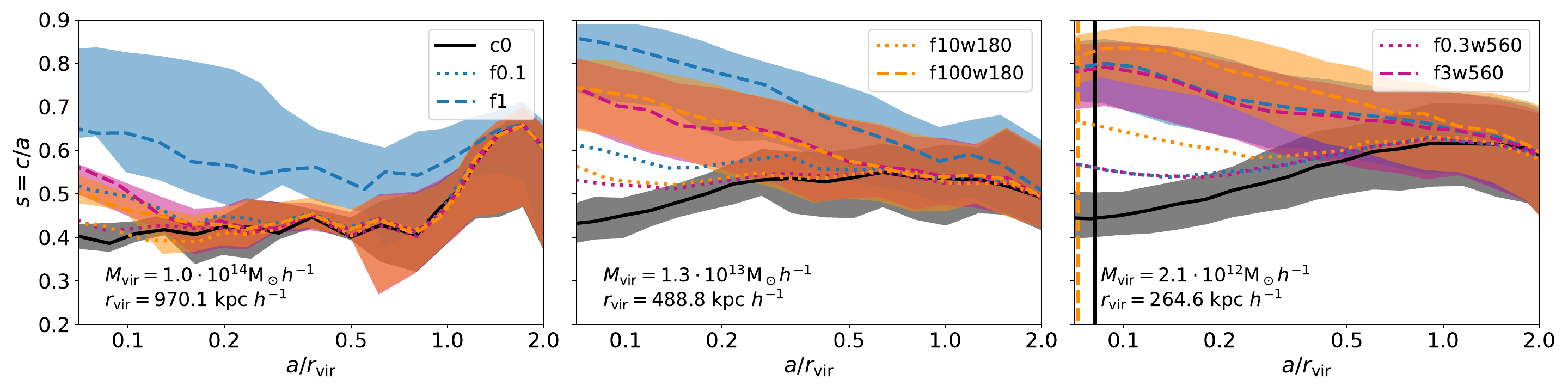}
    \caption{We show the median shape, $s=c/a$, of the DM haloes within three mass bins as a function of the major semiaxis, $a$.
    Each panel displays a different mass bin with its median mass being indicated.
    This figure is built analogously to the density profiles in Fig.~\ref{fig:density_profile}.
    The shaded regions indicate the scatter among the haloes, and the range between the 25th and 75th percentiles is displayed.
    We show it only for the collisionless DM and the strongest fSIDM model of each velocity-dependence.
    In addition, we indicate at which radius the shape sensitivity \citep[][]{Fischer_2023a} for the 25th percentile drops below a value of 25. This is indicative of a radius above which the shape measurements are reliable. Note, in particular for CDM the presence of satellites reduces the shape sensitivity.
    }
    \label{fig:shape_profile}
\end{figure*}

\begin{figure*}
    \centering
    \includegraphics[width=\textwidth]{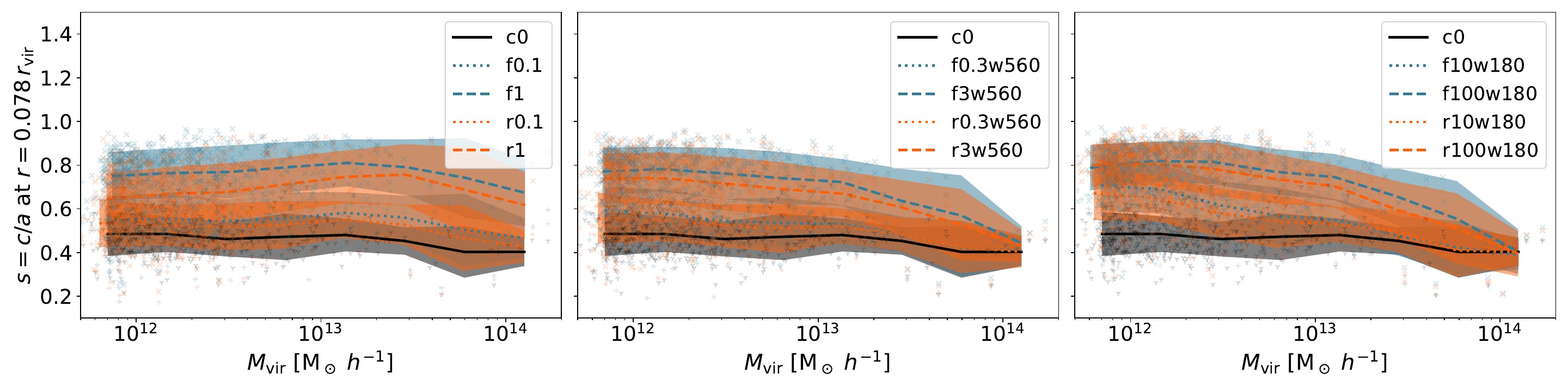}
    \caption{The shape of the DM haloes is shown as a function of their virial mass.
    The left-hand panel gives the results for the velocity-independent cross-sections \citep[previously shown in fig.~14 by][]{Fischer_2022}.
    In the middle panel, we display the results for the velocity-dependent scattering with $w=560 \, \mathrm{km}\,\mathrm{s}^{-1}$ and in the right-hand panel for $w=180 \, \mathrm{km}\,\mathrm{s}^{-1}$. This figure is built analogously to Fig.~\ref{fig:central_density}.
    }
    \label{fig:central_shape}
\end{figure*}

A commonly studied property of DM haloes is their shape. This has for SIDM been investigated by several authors \citep[e.g.][]{Peter_2013, Sameie_2018, Robertson_2019, Banerjee_2020, Chua_2020, Harvey_2021, Despali_2022, Shen_2022}.
DM self-interactions significantly affect the shape of the haloes up to larger radii than the density profile \citep{Fischer_2022}.
Furthermore, how large the affected radii are depends on the strength of the self-interactions \citep{Vargya_2022}.

To compute the shapes of our simulated DM haloes we proceed as previously described by \cite{Fischer_2022}.
We compute the mass tensor of particles within an ellipsoidal selection volume using their mass, $m$, and position, $r$:
\begin{equation} \label{eq:mass_tensor}
    \mathbf{M}_{ij} = \sum_k m_k r_{k,i} r_{k,j} \,.
\end{equation}
Here, $k$ denotes a particle and $i$, $j$ are the coordinate indices.
The selection volume for the next iteration is determined by the eigenvalues and eigenvectors of the mass tensor.
We iterate until the shape of the selection volume converges against the one inferred from the mass tensor.
It is important to note that shapes close to the centre of the haloes cannot be measured accurately.
The vanishing density gradient within the density core of SIDM haloes renders the shape undefined \citep[][]{Fischer_2023a}.

In Fig.~\ref{fig:shape_profile}, we plot $s=c/a$ as a function of the semimajor axis, $a$, in units of the virial radius.
The semiminor axis is denoted by $c$.
In general, we find that SIDM makes the haloes more round, as one would expect, and that fSIDM and rSIDM are qualitatively very similar.

Moreover, we show the shape of the haloes as a function of mass in Fig.~\ref{fig:central_shape}.
Here, we compute the shape from the innermost particles within a volume equal to a sphere of radius $0.078 r_\mathrm{vir}$.
For CDM, we find that haloes become more ellipsoidal with increasing mass. This trend is well known in the literature \citep[e.g.][]{Jing_2002, Allgood_2006, Munoz-Cuartas_2011, Despali_2013, Despali_2014}.
This can change when including self-interactions, especially for a velocity-independent cross-section.
Here, the effect of the self-interactions is increasing with halo mass (see the left-hand panel of Fig.~\ref{fig:central_shape}).
However, for the most massive systems in our simulation we find the haloes to become more elliptical even with SIDM.
This might be due to few objects which on average might be less relaxed than the ones at lower masses.
Given a velocity dependent cross-section haloes become more elliptical with mass at the high-mass end.
But the gradient is steeper compared to CDM as self-interactions lead to rounder haloes at lower masses and at the high-mass end the shape becomes similar to CDM (middle and right-hand panel).

Overall, we reproduce the same trends as in previous SIDM simulations.
As far as we can compare, our results are in broad agreement with the shapes reported in other studies \citep[e.g.][]{Peter_2013}.

\subsubsection{Satellites} \label{sec:results_satellites}

\begin{figure*}
    \centering
    \includegraphics[width=\textwidth]{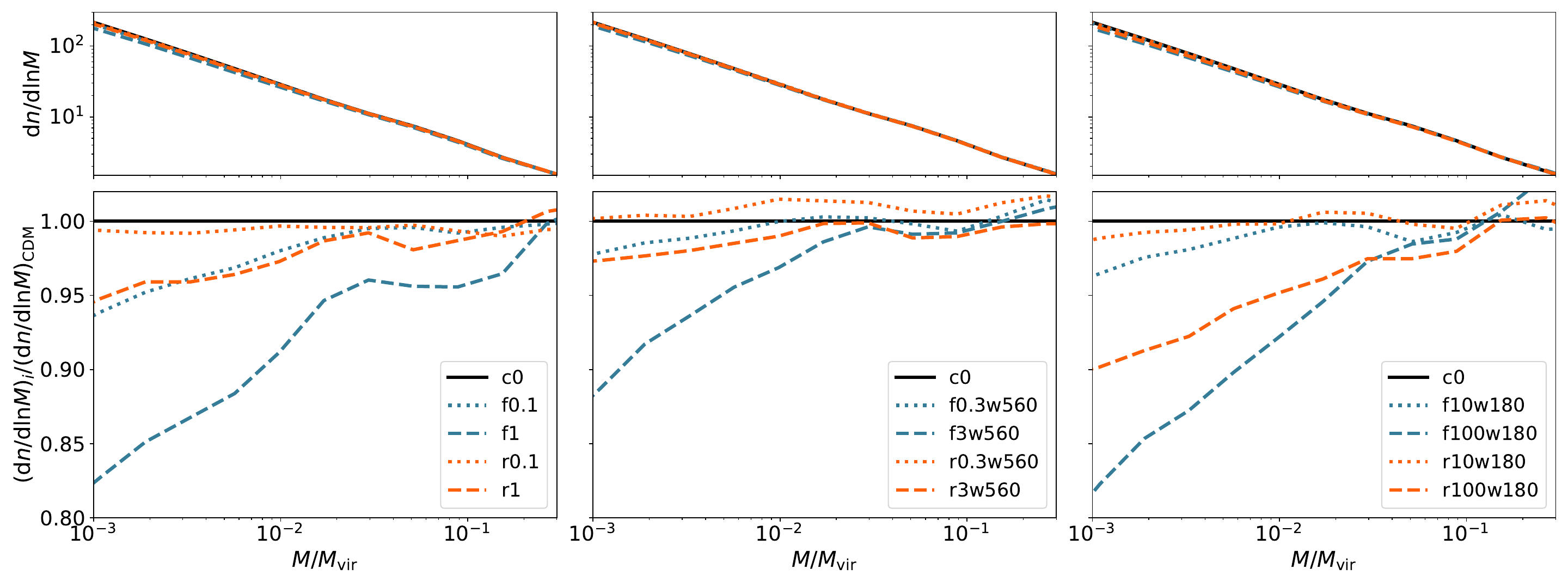}
    \caption{We show the number of satellites per logarithmic mass as a function of their total mass relative to the virial mass of their host (upper panels). In the lower panels, we display the ratio of the DM models to CDM.
    All panels give the result of the 100 most massive groups in our full cosmological box.
    The left-hand panels show the results for the velocity-independent cross-sections \citep[previously shown in fig.~6 of][]{Fischer_2022}.
    The middle panels gives the velocity-dependent self-interactions with $w = 560 \, \mathrm{km}\,\mathrm{s}^{-1}$ and the right-hand side panels for $w = 180 \, \mathrm{km}\,\mathrm{s}^{-1}$.
    All subhaloes, except for the primary one, within a radius of $5 \, r_\mathrm{vir}$ were considered.
    The results are for a redshift of $z = 0$.
    Note that the least resolved satellites used here contain about 100 particles.}
    \label{fig:shmf}
\end{figure*}

\begin{figure*}
    \centering
    \includegraphics[width=\textwidth]{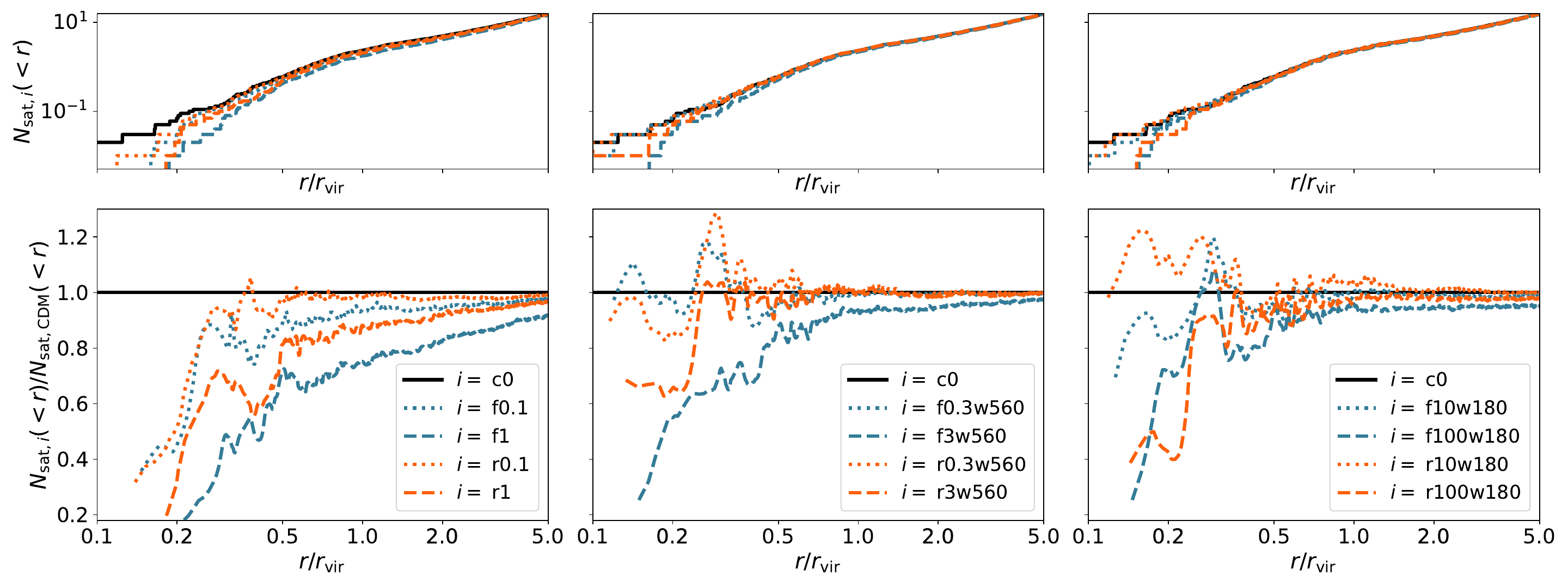}
    \caption{For the 100 most massive haloes of our simulations, we show the cumulative number of satellites per halo as a function of radius (upper panels).
    We also give the ratio of the DM models to CDM (lower panels).
    The left-hand panel shows the results for the velocity-independent cross-sections \citep[previously shown in fig.~7 of][]{Fischer_2022}.
    The middle panel gives the velocity-dependent self-interactions with $w = 560 \, \mathrm{km}\,\mathrm{s}^{-1}$ and the right-hand side panel for $w = 180 \, \mathrm{km}\,\mathrm{s}^{-1}$.
    The results are shown for $z=0$ and subhaloes were only considered if they are less massive than the primary subhalo and more massive than $M > 9.6 \times 10^{10} \, \mathrm{M_\odot} \, h^{-1}$.}
    \label{fig:satellites}
\end{figure*}

The properties of satellite systems are a promising probe for studies of DM.
Depending on the DM model, fewer or more satellites are predicted, and they may differ in their density profiles.
This has been studied in the context of multiple DM models, including SIDM \citep[e.g.][]{Banerjee_2020, Nadler_2020, Nadler_2021, Bhattacharyya_2022}.

In Fig.~\ref{fig:shmf}, we show the number of satellites per logarithmic mass as a function of their mass in units of the virial mass of their host system.
We find that DM self-interactions can reduce the abundance of satellites, and the number of less massive subhaloes is stronger affected than the more massive satellites.
Moreover, the momentum-transfer-matched frequent self-interactions lead to a stronger suppression than the isotropic scattering \citep[as previously described for a constant cross-section in][]{Fischer_2022}.
All this seems to be independent of the velocity dependence.
Interestingly, the difference between fSIDM and rSIDM is shrinking for the strong velocity dependence.
For the velocity-independent simulations (left-hand panel) and the mildly velocity-dependent runs ($w=560\,\mathrm{km}\,\mathrm{s}^{-1}$, middle panel), the stronger rSIDM cross-section has a similar effect to the weak fSIDM cross-section.
But for the strongly velocity-dependent run ($w=180\,\mathrm{km}\,\mathrm{s}^{-1}$, right-hand panel), the strong rSIDM cross-section is no longer similar to the weak fSIDM one but closer to the strong fSIDM one.
Hence, we find a strong velocity dependence to reduce the differences between cross-sections with different angular dependencies.

The difference between rSIDM and fSIDM may mainly arise from host-satellite scattering as those interactions take place with a preferred direction and thus are far from an equilibrium state.
Also, these interactions significantly contribute to the suppression of the satellite abundance \citep[e.g.][]{Zeng_2022}.
To understand the reduced difference between rSIDM and fSIDM, it is important to note that the host-satellite interactions take place at higher velocities than the scatterings within the satellite between its particles. Consequently, a velocity-dependent cross-section can reduce the host–satellite scattering compared to the satellite-satellite interactions and thus reduce the difference between rSIDM and fSIDM.

\begin{figure*}
    \centering
    \includegraphics[width=\textwidth]{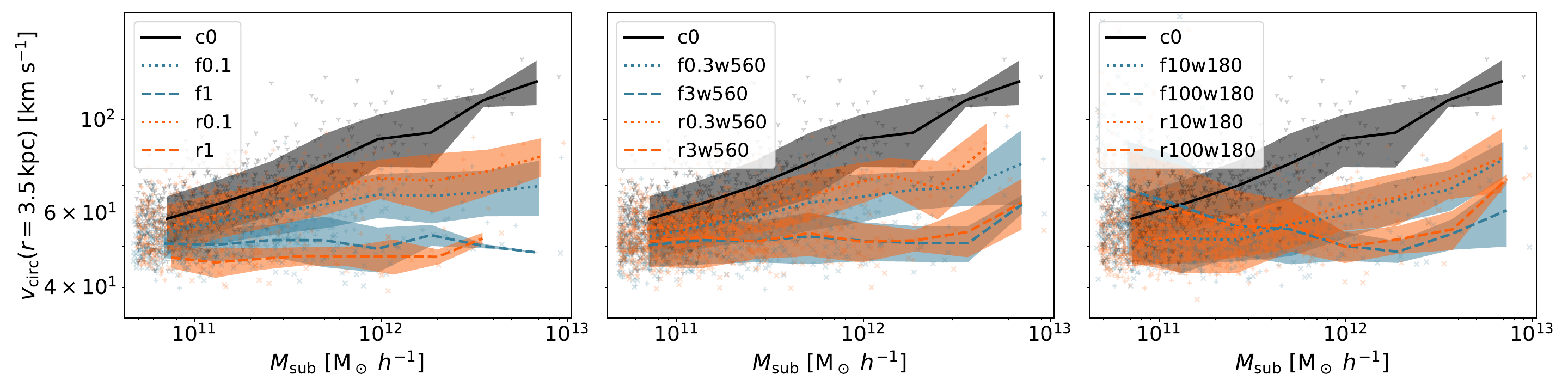}
    \caption{We show the circular velocity at $3.5 \, \mathrm{kpc}$ for satellites with a mass of at least $\approx 4.9 \times 10^{10}\, \mathrm{M_\odot} \, h^{-1}$. We consider all satellites that are not the primary subhalo.
    The lines indicate the mean and the shaded regions the standard deviation for the corresponding DM models.
    This is analogous to Fig.~\ref{fig:central_density}, as well as the markers.
    }
    \label{fig:vel_circ}
\end{figure*}

In addition, we find that the suppression of the satellite abundance for the mildly velocity-dependent cross-sections (middle panel) is less strong than for the other two velocity dependencies.
We would not have expected this difference in strength from the density profiles that we show in Sec.~\ref{sec:results_density_profile}.
Though, there is a velocity scale at which the mildly velocity-dependent cross-sections are weaker than the corresponding ones with a different velocity-dependence (see Fig.~\ref{fig:cross-sections}).
Interestingly, this becomes even more pronounced when computing the effective cross-section introduced by \citet[][see Appendix~\ref{sec:constraints}]{Yang_2022D}.
Given that the host-satellite scattering, which drives the suppression of the satellite abundance, takes preferentially place in this velocity regime, it could explain the different strengths of the satellite suppression.

In Fig.~\ref{fig:satellites}, we display the number of satellites as a function of the distance to their host in units of the host's virial radius.
The upper panels show the cumulative number of satellites and the lower panels display the ratio to CDM. 
We note that the ratios at small distances are subject to a considerable amount of noise as they are computed from a small number of satellites.
Here, we find again that self-interactions can suppress the number of satellites.
The inner ones are more affected than the distant ones and frequent self-interactions lead to a stronger suppression than rare scattering if the same momentum transfer cross-sections are compared.
This is well visible for the velocity-independent cross-sections in the left-hand panel.
The simulations with frequent self-interactions show roughly a reduction in the number of satellites twice as large as for the corresponding simulations with rare self-interactions.
As in Fig.~\ref{fig:shmf}, we find that the difference between rSIDM and fSIDM becomes less for the strongest velocity-dependence ($w = 180\,\mathrm{km}\,\mathrm{s}^{-1}$).

\subsubsection{Diversity of satellites} \label{sec:results_diversity}

One of the small-scale issues is the diversity problem.
It usually refers to the variation between the rotation curves of galaxies \citep[e.g.][]{Kamada_2017, Ren_2019, Zentner_2022}.
To study their diversity, we focus on the circular velocity at a radius of $3.5 \, \mathrm{kpc}$ instead of looking at the full profile.
The velocity at $3.5 \, \mathrm{kpc}$ is sensitive to the core formation or core collapse.
In Fig.~\ref{fig:vel_circ}, we show the circular velocity at that radius for satellites more massive than $\approx 4.9 \times 10^{10}\, \mathrm{M_\odot} \, h^{-1}$ as a function of their mass.
Note that we consider all subhaloes identified by \textsc{SubFind} satellites if they are not a primary subhalo (see Sec.~\ref{sec:simulations}).

For the velocity-independent cross-sections (the left-hand panel of Fig.~\ref{fig:vel_circ}), we find that self-interactions decrease the circular velocity at $3.5 \, \mathrm{kpc}$.
This corresponds to the formation of a density core.
For the larger cross-sections the circular velocity is lower, i.e.\ the density core is larger.
Basically, the same applies for the cross-sections with $w=560 \, \mathrm{km} \, \mathrm{s}^{-1}$ (the middle panel of Fig.~\ref{fig:vel_circ}).
But it is noticeable that the most massive subhaloes experience less suppression of $v_\mathrm{circ}$ in the inner region.
This is simply a consequence of the velocity dependence, as the DM particles in the more massive subhaloes have higher typical relative velocities.
For the cross-section with the strong velocity dependence ($w=180 \, \mathrm{km} \, \mathrm{s}^{-1}$), we find qualitatively different results.
For the more massive subhaloes, we find the suppression of the circular velocity as for the other simulations.
But on average, the least massive objects show an increase in circular velocity for the stronger cross-sections compared to CDM.
The satellites with larger circular velocities are more compact, i.e.\ they contain more mass within $r=3.5 \, \mathrm{kpc}$.
Moreover, we also found that the inner density gradients are steeper (see Appendix~\ref{sec:density_gradient}).
This is an indication that they have entered the collapse phase.
Moreover, the distribution of values for the circular velocity is broader at low masses compared to CDM.
The other cross-sections do not show such a significant increase in diversity.

When comparing the results for rSIDM and fSIDM, we do not find a clear qualitative difference arising from the typical scattering angle of the self-interactions.
In contrast, the momentum transfer cross-section provides a matching that is not far off but surprisingly accurate for the velocity-dependent cross-sections.

The diversity of rotation curves has been studied a lot with SIDM, and it has been shown that self-interactions can create more diverse density profiles.
In particular, low-mass objects have been studied.
There are several papers that studied MW-like satellites and dwarf galaxies \citep[e.g.][]{Creasey_2017, Zavala_2019, Correa_2022, Lovell_2023}.
It has been found in DMO simulations that cross-sections with a strong velocity dependence can even trigger core collapse within satellites \citep[e.g.][]{Turner_2021, Yang_2023Da, Nadler_2023}.
Especially for satellites the core collapse can be enhanced by tidal stripping \citep[e.g.][]{Kahlhoefer_2019,Nishikawa_2020}.
This is in line with our finding of more compact objects at low masses for our strongly velocity-dependent cross-sections.

\subsubsection{Frequent versus rare self-interactions} \label{sec:results_frequent_vs_rare}

\begin{figure*}
    \centering
    \includegraphics[width=0.65\columnwidth]{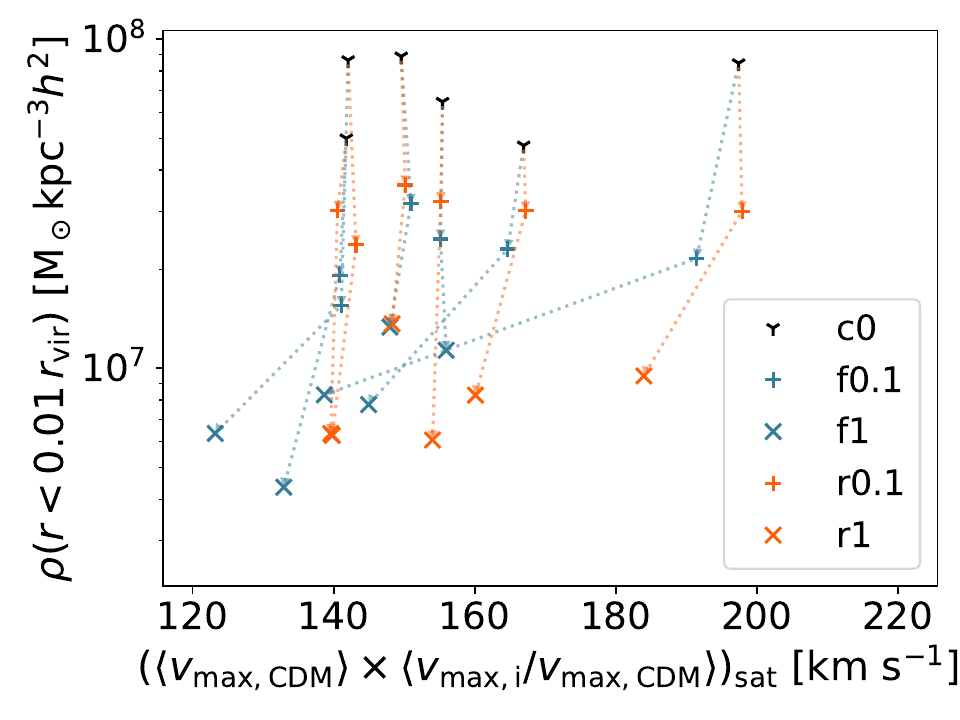}
    \includegraphics[width=0.65\columnwidth]{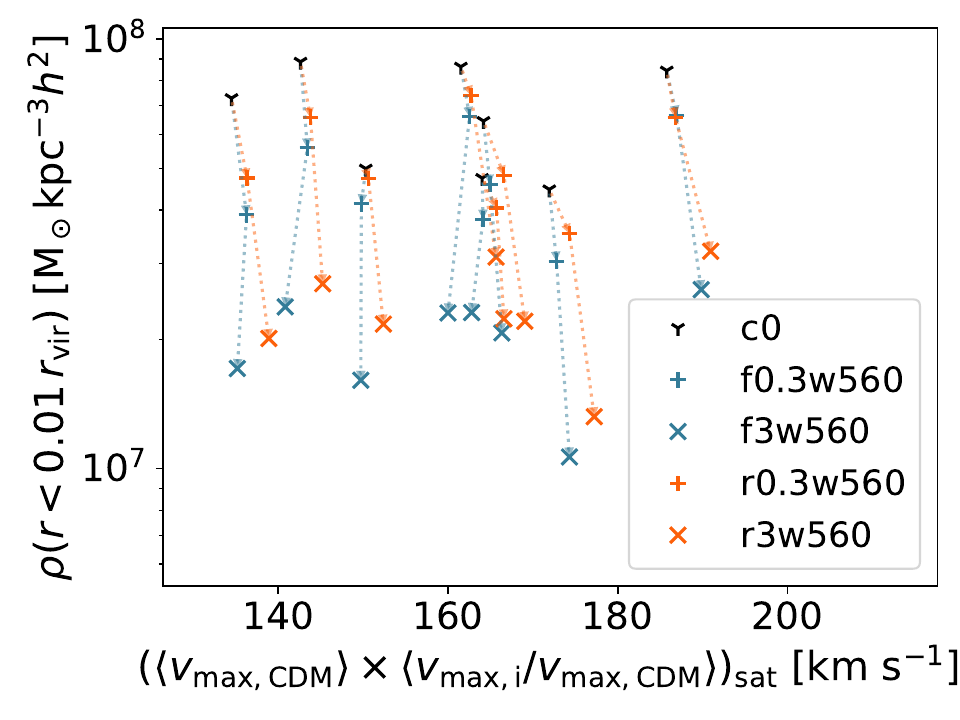}
    \includegraphics[width=0.65\columnwidth]{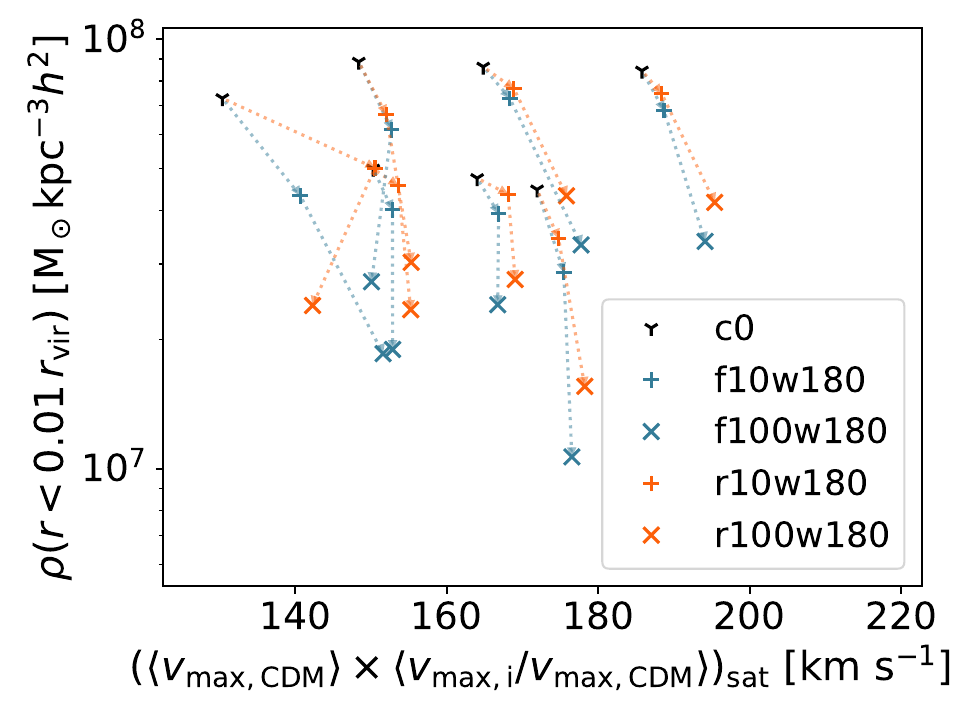}
    \caption{We show how the DM model affects the maximum circular velocity in the satellites and the host’s central density. We have cross-identified the haloes in the different DM runs. The lines connect the same halo, i.e.\ indicated how the properties of a halo change when varying the cross-section. The shown haloes are among the most massive ones, the details of the selection criterion are explained in the text.}
    \label{fig:frequent_vs_rare}
\end{figure*}

Finally, we want to investigate how the different DM models affect the satellites of our most massive haloes.
Previously, we found that fSIDM can lead to a stronger suppression of the number of satellites than rSIDM does \citep{Fischer_2022}.
Identifying such differences is crucial to constrain the angular dependence of DM self-interactions.
In contrast to our previous work, we investigate the maximum circular velocity in the satellites here.
But show the number of satellites in Appendix~\ref{sec:results_frequent_vs_rare_add}.

We cross-identify the haloes and their satellites among the simulations based on their particles.
As we start from the same initial conditions, we can match the haloes with the same particles identified based on their unique identification numbers.
To evaluate how well two haloes match we make use of the gravitational potential at the particle's location.
Particles at a lower gravitational potential are stronger weighted to find the best matching analogue.
Given a list of the halo particles sorted according to how deep they sit in the gravitational potential, starting with the one at the lowest potential, we compute weights for them. These weights are given as
\begin{equation}
    w_i = \left(\frac{1}{i+1}\right)^\alpha \,.
\end{equation}
Note, here we assume the first list index to be $i=0$.
The parameter $\alpha$ allows for different weightings, we use $\alpha=0.8$.
In practice, we compute the weight for the CDM run only.
This is because we use the CDM haloes as a benchmark and ask how well the SIDM haloes match them.
The quality of a potential match is given by the sum of the weights $w_i$ for the particles that the CDM halo and the SIDM halo have in common.

For the analysis, we do not consider all haloes but apply different selection criteria.
Firstly, the hosts and their satellites should be well-resolved.
We consider only the 13th most massive haloes and limit the selection further by requiring that we are able to match at least five satellites with a minimum mass of $9.6 \times 10^{10}\, \mathrm{M_\odot} \, h^{-1}$ (2200 particles).
Furthermore, we require the haloes to be relaxed.
Here, we assume a halo to be relaxed if the centre of mass and the most bound particle of the primary subhalo are separated by not more than $10\%$ of the virial radius.
In addition, we tested a further limitation by excluding haloes based on the ratio of the halo and primary subhalo mass.
However, in practice, this did not exclude any halo.
At least when we have required that the primary subhalo does not contain less than $75\%$ of the halo mass.

In Fig.~\ref{fig:frequent_vs_rare}, we display our results for how the central halo densities correlate with the relative change of the maximum circular velocity in the satellites.
We show the average relative change multiplied by the average maximum circular velocity in the CDM satellites.
Here, we use the maximum velocity as computed by \textsc{SubFind}.
It is given by the maximum of the circular velocity, $v_\mathrm{circ} = \sqrt{\mathrm{G}\,M(<r) / r}$, in radial distance, $r$, from the centre of the subhalo.

We find the maximum circular velocity in the satellites altered by the DM self-interactions.
For the velocity-independent scattering it typically decreases with increasing cross-section.
This implies that the satellites are less concentrated.
In contrast, a velocity-dependent cross-section can also lead to a larger value for the maximum circular velocity.
Whether this is the case or not depends in our model on the parameter $w$, i.e.\ how strongly velocity-dependent the scattering is.
It is worth pointing out that our selection criterion of subhaloes above a mass threshold that we can match might favourably pick subhaloes that have become more concentrated due to the velocity-dependent self-interactions.
Thus, the increase in maximum circular velocity may not be representative of all the subhaloes.

We find that frequent self-interactions tend to lead to a smaller maximum circular velocity than rare scattering.
For the larger cross-sections we have simulated, we find that the maximum circular velocity for rare self-interactions compared to frequent ones is increased for the typical system (median) by $\approx8 \%$ (velocity-independent), $\approx2 \%$ ($w=560\,\mathrm{km}\,\mathrm{s}^{-1}$), and $\approx1 \%$ ($w=180\,\mathrm{km}\,\mathrm{s}^{-1}$).
This means that the difference between fSIDM and rSIDM decreases for our simulations with stronger velocity dependence.
Hence, this is in line with our finding of a qualitative difference for the abundance of satellites in Section~\ref{sec:results_satellites}.
However, the difference we find here might also largely be due to the fact that the stronger velocity-dependent cross-section we study has a weaker effect on massive haloes.
For example, this becomes visible when comparing the central densities.
In consequence, the reduced qualitative difference between large- and small-angle scattering might be better visible from Fig.~\ref{fig:shmf}.
But here we can see that not only for a constant cross-section the angular dependence matters but also for strongly velocity-dependent self-interactions even if the subhaloes are becoming more compact on average.

We note that the analysis above is not based on a larger statistical sample and thus the exact numbers may change.
But we expect the qualitative trend to be the same.
It is also worth pointing out that the less massive satellites might be affected more strongly by the self-interactions (see Fig.~\ref{fig:shmf}) and thus differences between models are larger for them.
Hence, this should be followed up with simulations with a much higher spatial resolution.


\section{Discussion} \label{sec:discussion}
In this section, we discuss the assumptions and limitations of our simulations as well as the implications of our results. 
We begin with technical considerations and end by discussing what the next steps for a follow-up study may look like.

In contrast to our previous work \citep[][]{Fischer_2022}, we explored velocity-dependent cross-sections.
We found that simulating those interactions requires a separate time-step criterion \citep[i.e. different from the one of][]{Fischer_2021b}.
Especially cross-section with a strong velocity-dependence, i.e.\ a small value for $w$ (see equation~\ref{eq:veldep}), can be computationally very expensive compared to a velocity-independent cross-section with a similar effective cross-section. A more detailed discussion of building a time-step criterion can be found in Appendix~\ref{sec:time_step_disc}.

When measuring the core sizes in Sec.~\ref{sec:results_isolated_haloes}, we found that the resulting fit is surprisingly sensitive to the optimization method.
This may limit the comparability of core sizes inferred by different authors.
In particular, \cite{Correa_2022} describe in their appendix~B, that results on the evolution of the core size differ in terms of the maximum cores size in the literature.

The results of our cosmological simulations depend on the algorithms employed to identify haloes and their substructure.
For this task, we used the build-in module \textsc{SubFind} \citep[][]{Springel_2001, Dolag_2009}.
There exist a number of codes that are capable of identifying substructure \citep[e.g.][]{Knollmann_2009, Maciejewski_2009, Tweet_2009, Behroozi_2012, Han_2017, Elahi_2019}.
These codes use different algorithms and are known to give somewhat different results \cite{Knebe_2013}.
In consequence, our results could change a bit when employing a different substructure finder.

In this paper, we aimed to understand how a velocity dependence of the self-interactions affects differences arising from the angular dependence of the cross-section.
Very anisotropic cross-sections are typically expected to be velocity-dependent \citep[e.g.][]{Buckley_2010, Loeb_2011, Bringmann_2017}.
It is known that fSIDM and rSIDM differ mainly in systems that are far from equilibrium, such as mergers \citep{Fischer_2021a} and the abundance of satellites \cite{Fischer_2022}.
The evolution of those systems is governed by multiple velocity scales, where typically the larger velocity scale is the one that is mainly responsible for differences arising from the angular dependence of the self-interactions.
Consequently, the difference becomes less when the self-interactions at large velocities are suppressed due to velocity-dependent scattering.
We found this for the abundance of satellites.
In consequence, it could be interesting to probe less massive systems for distinguishing rSIDM and fSIDM as the velocity dependence could be weaker.
At least in the model employed in our study, a system with typical velocities smaller than $w$ would only experience a weak velocity dependence (see equation~\ref{eq:veldep}).
The relevant mass scales for the cross-sections we simulated are visible from the effective cross-section as a function of mass shown in Appendix~\ref{sec:constraints}.

Despite our studies of satellites, it is worth mentioning that very anisotropic cross-sections have been mainly studied in the context of merging galaxy clusters \citep[e.g.][]{Kahlhoefer_2014, Harvey_2015, Fischer_2023b, Wittman_2023}.
At about the pericentre passage, such cross-sections can give rise to an effective drag force decelerating the DM component and creating an offset between the galaxies and the DM.
Cross-sections that are velocity dependent and strongly anisotropic, have not been studied in the context of such mergers yet.
Only a Bullet Cluster-like system has been simulated by \cite{Robertson_2017b} using a velocity-dependent anisotropic cross-section, but it does not fall within the limit of fSIDM.
Studying merging systems with velocity-dependent fSIDM is crucial to understand their power to constrain such models and is the subject of a companion paper \citep{Sabarish_2023}.

Our simulations are all DM-only.
On the one hand, it allows us to understand the qualitative differences between DM models better compared to simulations including further physical processes.
But on the other hand, it limits the possibility to compare the results to observations and derive constraints on the cross-section.
Consequently, the next step would be to include baryonic physics, i.e.\ run hydrodynamical simulations.
Several authors have found that taking baryons into account can reduce the differences between collisonless and self-interacting DM and thus would mitigates constraints derived from DM-only studies \cite[e.g.][]{Fry_2015, Despali_2022, Sirks_2022, Mastromarino_2023}.
SIDM can be more responsive to the baryon distribution than CDM in Milky Way-mass galaxies \citep[e.g.][]{Sameie_2018, Sameie_2021}.
In the presence of baryons, effects from SIDM can even be reversed – at least for a fraction of the haloes.
It has been shown that for galaxies with Milky Way-like masses and above the interplay of baryons and self-interactions can lead to cuspier density profiles than in CDM \citep[e.g.][]{Despali_2019, Rose_2022}.
In principle, baryons could also affect the ability to constrain the angular dependence with the abundance of satellites.

Aside from constraining the angular dependence, one would like to have a procedure to compare the effect of SIDM with different angular dependencies.
This would allow to transfer constraints between models that differ in their typical scattering angle.
\cite{Yang_2022D} introduced the effective cross-section for this purpose, where the angular matching is based on the viscosity cross-section.
However, the quality of the matching may depend on the physical system, i.e.\ how relaxed the system is.
But not only on this, as we found the momentum transfer cross-section can at least for some set-ups provide an excellent match (see Fig.~\ref{fig:nfw_core}) excluding that the viscosity cross-section does as well.
However, this does not contradict the viscosity cross-section providing a better match usually.
But it implies that the matching is more complicated and may depend on the properties of the astrophysical system.
It may matter how strong the self-interactions are and whether the system evolves in the smfp or lmfp regime.
In the latter, one gravity plays an important role between two consecutive scattering events (assuming an isotropic cross-section) and thus may make the evolution of the halo and the matching of different angular dependencies sensitive to the details of the density profile.


\section{Conclusions} \label{sec:conclusions}

In this paper, we have studied SIDM with velocity-dependent scattering, considering isotropic cross-sections and strongly forward-enhanced ones.
For accurate modelling of velocity-dependent self-interactions, we introduced a new time-step criterion and enhanced the performance with an improved parallelization scheme.
To learn about qualitative differences arising from the velocity dependence, we first simulated the thermalization problem, a simple test problem without gravity. 
Secondly, we studied the evolution of the density profile of isolated haloes including Hernquist and NFW profiles.
For the remainder of the paper, we focused on cosmological simulations and investigated the qualitative differences between the DM models concerning the velocity and angular dependence of the self-interactions.
Our most important results can be summarized as follows:

\begin{itemize}
    \item We found that velocity-dependent self-interactions lead to a slower population of the high-velocity tail of the Maxwell-Boltzmann distribution during thermalization due to the suppressed cross-section at high velocities.
    
    \item The evolution of the density profile of isolated haloes is qualitatively affected by the velocity dependence, i.e.\ it is not self-similar. This can lead to a longer collapse time relative to the core formation time and a larger maximum core size. However, we found a significant difference between velocity-independent and velocity-dependent cross-sections only for strong velocity dependencies, i.e.\ when $w$ is much smaller than the typical scattering velocity.
    
    \item The velocity dependence of the self-interactions controls whether the central density of haloes is increasing or decreasing as a function of halo mass.
    
    \item Given a strong velocity dependence (small $w$), frequent self-interactions can diversify the density profile similar to an isotropic cross-section. We found that the two angular dependencies can create haloes that are less compact as well as haloes that are more compact at the same subhalo mass. This makes SIDM, regardless of its angular dependence, promising to explain the observed diversity.
    
    \item A strong velocity dependence of the cross-section, i.e.\ a small value of $w$ can reduce the differences between fSIDM and rSIDM regarding the abundance of satellites.
\end{itemize}

The simulations we conducted were DM-only and allowed us to understand phenomenological differences arising from the velocity dependence of DM scattering.
Our results may be instructive for more detailed studies of qualitative differences between SIDM models and helpful in designing more sophisticated simulations that include baryonic matter and additional physics such as cooling, star formation, AGN, and associated feedback mechanisms.
Undertaking such a study to learn about the chances to discriminate between rSIDM and fSIDM when baryonic physics is taken into account, is the subject of forthcoming work.


\section*{Acknowledgements}
We want to thank the anonymous referee for helpful comments that improved the paper.
In addition, we would like to thank all participants of the Darkium SIDM Journal Club for helpful discussions.
MSF thanks Lucas Kimmig for a fruitful discussion.
This work is funded by the Deutsche Forschungsgemeinschaft (DFG, German Research Foundation) under Germany's Excellence Strategy – EXC 2121 ``Quantum Universe'' –  390833306, Germany’s Excellence Strategy – EXC-2094 ``Origins'' – 390783311 and the Emmy Noether Grant No.\ KA 4662/1-2.
Antonio Ragagnin acknowledge support from the grant PRIN-MIUR 2017 WSCC32.
KD acknowledges support by the COMPLEX project from the European Research Council (ERC) under the European Union’s Horizon 2020 research and innovation programme grant agreement ERC-2019-AdG 882679.
The simulations have been carried out on the computing facility Hummel (HPC-Cluster 2015) of the University of Hamburg and the computing facilities of the Computational Center for Particle and Astrophysics (C2PAP). Preprint numbers: DESY-23-154, TTP23-041.

Software:
NumPy \citep{NumPy},
Matplotlib \citep{Matplotlib},
SciPy \citep{SciPy}

\section*{Data Availability}

The data underlying this article will be shared on reasonable request to the corresponding author.
In addition, some of the data can be retrieved from our webpage: \url{https://www.darkium.org}.



\bibliographystyle{mnras}
\bibliography{references}




\appendix

\section{Parallelization} \label{sec:parallelisation}

The improved MPI parallelization orders communication with the aim to reduce waiting time.
Therefore, a separate communication list is computed at the cost of additional overhead.

The communication list is computed in seven steps:
\begin{enumerate}
    \item Every MPI rank already knows how many particles it will receive from and send to other processes.
    It creates a list with all its communications to other processes; here, the sum of the particles that are exchanged with each process (sum of send and receive) is stored. This number will later be used to assign individual communications a priority.
    \item Each process sorts the previously created list according to priority.
    \item All processes exchange their communications list. In turn, every process has all communication lists. We note that the length of these lists does scale with less than $N^2$ ($N$ being the number of MPI ranks) if no communication between all processes is required (this is typically the case).
    \item Every process builds a single list with all communications. Starting with the highest priority elements of all individual lists.
    \item The list from the previous step is sorted according to priority.
    \item The overall communication plan is built by every process to avoid additional communication.
    It contains several communication steps. At each step, several pairs do their communication and also the scattering computations of the exchanged particles. One process can be assigned only one pairwise communication per step.
    The plan is built by trying to fulfil the highest priorities first. We start with the first communication step and try to fill in the communications sorted after priority. If a communication does not fit in a step because a corresponding process is already busy, it is queued and retried for the next step. We turn to the next step when the queue for the communication that should be assigned for the current step is empty or when all processes are already assigned a communication for the current step.
    \item Each process extracts its communication schedule from the overall communication plan.
\end{enumerate}

For each pair of processes we consider, the first sends the particles and the second receives them and does the computation.
When the computation has finished, the particles are sent back.
Subsequently, the communication is done in the other direction, i.e.\ the other task will compute the scattering of the particles.

\begin{figure}
    \centering
    \includegraphics[width=\columnwidth]{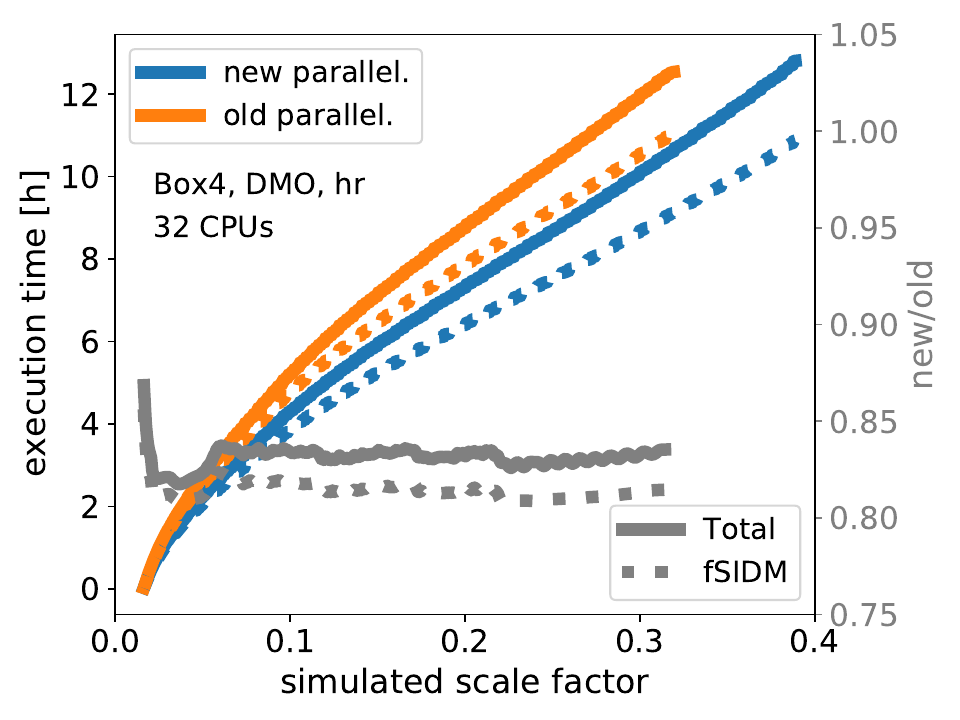}
    \includegraphics[width=\columnwidth]{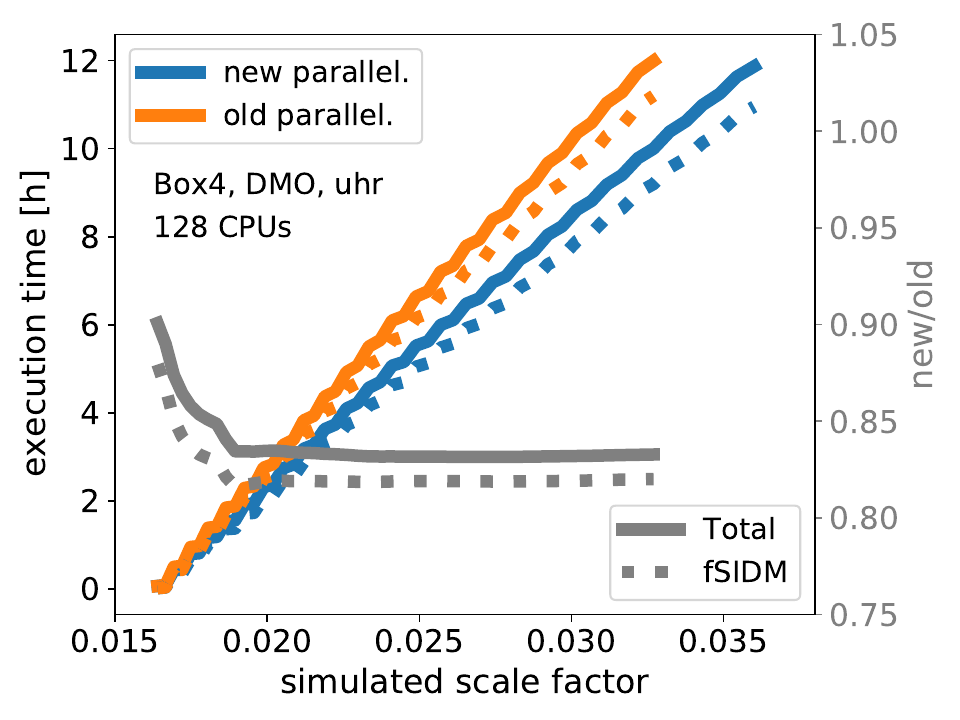}
    \caption{We show the execution time as a function of simulated time for the Magneticum Box 4 with high resolution (hr) and ultra-high resolution (uhr).
    The simulations are DMO.
    We also show the time spend on the fSIDM-related computations, which make up the majority of the computational costs.
    Note, that we do not use adaptive gravitational softening. Consequently, the computation of the kernel sizes is counted as fSIDM-related.
    In grey, we show the ratio of execution time between the old and new parallelization.
    }
    \label{fig:perform_parallel}
\end{figure}

We have run a performance test for Box 4 of the Magneticum simulations with DM only using the high-resolution (hr) and ultra-high-resolution (uhr) initial conditions \citep[the same as in][]{Fischer_2022}.
The results using 32 and 128 MPI ranks are displayed in Fig.~\ref{fig:perform_parallel}.
For this test, we used only the MPI parallelization and did not make use of an OpenMP parallelization that exists for other parts of the code.
Clearly, the improved parallelization leads to a significant speed-up.
It is visible that we save with the improvements more than $15\%$ of the computational costs
This parallelization has been used for some of the simulations described in this paper.

\section{Time-step criterion discussion} \label{sec:time_step_disc}
In the following, we discuss the thoughts behind the construction of a time-step criterion for SIDM in further detail. Here, we aim to build a time-step criterion that ensures that all or almost all interactions take place at a sufficiently small time-step. This differs from the approach taken by \cite{Vogelsberger_2012}, which considered the local velocity dispersion. Instead, we are concerned about the full velocity distribution, i.e.\ which relative velocities a particle actually sees.

Previously we introduced a time-step criterion for velocity-independent self-interactions \citep{Fischer_2021b}. That time-step criterion estimates the time-step based on the maximum velocity that a particle experienced in the previous time-step\footnote{Strictly speaking, we use $\sigma_{\mathrm{T}}/m \, \Delta v$, but for a velocity-independent cross-section $\sigma_{\mathrm{T}}/m$ is a constant and thus the same for all interactions.} and the maximal possible kernel overlap, $\Lambda_{ii}$\footnote{Actually, we did not directly compute $\Lambda_{ii}$, but used $h_i^{-3}$ as an estimate ($h$ denotes the kernel size).}. Given that the neighbour number, $N_\mathrm{ngb}$, is sufficiently large, the velocities that a particle has seen in the previous time-step allow us to roughly estimate the maximum velocity it may experience in the next time-step.
In contrast, for a velocity-dependent cross-section, the relevant velocity is not the maximal velocity that a particle may see but how close it gets to the velocity for which the interaction probability, or generally speaking, $\sigma_{\mathrm{T}}/m \, \Delta v$, becomes maximal. Actually, this is also what the time-step criterion by \cite{Fischer_2021b} tries to estimate when used for a velocity-dependent cross-section. But in contrast to a constant cross-section, it is much harder to estimate this for a velocity-dependent cross-section based on the velocities a particle has seen, as the probability of seeing a relative velocity close to $v_e$ (see equation~\ref{eq:time_step_ve}) might be small. Our tests showed that we would often overestimate the time-step.
This problem can be circumvented by directly using $v_e$, instead of making estimates based on what a particle has seen in the previous time-step as described in Sec.~\ref{sec:time_step_crit}. Hence, it is possible to build a timestep criterion that guarantees for each particle pair the interaction probability or drag force is sufficiently small.

Lastly, we want to explain why one should not directly use the interaction probabilities a particle has encountered in the previous time-step. The disadvantage is that large interaction probabilities are less likely to be seen by a numerical particle than the relevant velocities (for a constant cross-section this would be the maximum velocity). This is because in most cases the kernel overlap, $\Lambda_{ij}$, is small. Or in other words, the probability of having a numerical particle pair with a relative velocity close to the relevant velocity and a large kernel overlap is smaller than having only a relative velocity close to the relevant velocity.

\section{Comoving Integration Test} \label{sec:comoving_integration_test}

\begin{figure}
    \centering
    \includegraphics[width=\columnwidth]{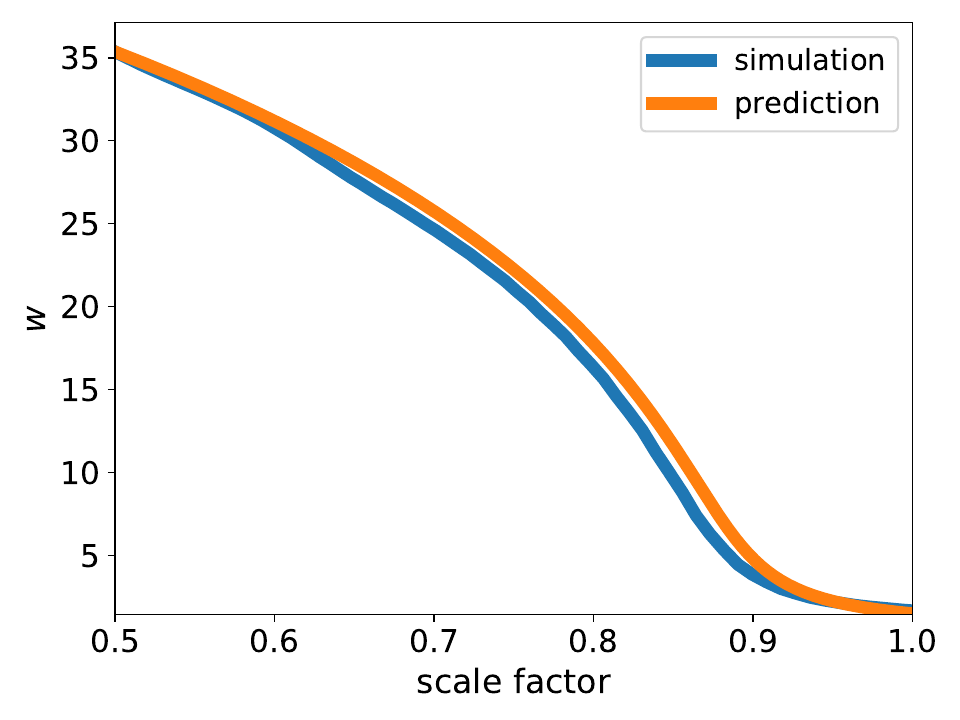}
    \caption{The cosmic deceleration problem in terms of canonical momentum is shown.
    The simulation runs from $a=0.5$ to $a=1.0$ with $122500$ particles in a cubic box with a comoving side length of $1400 \, \mathrm{kpc} \, h^{-1}$.
    The total mass is $22.8465 \times 10^{10} \, \mathrm{M_\odot} \, h^{-1}$, corresponding to a comoving density of $83.26 \, \mathrm{M_\odot}\,\mathrm{kpc}^{-3}\,h^2$.
    The initial snapshot velocity of the test particle is $100 \, \mathrm{kpc}\, \mathrm{Gyr}^{-1}$, which corresponds to an initial canonical momentum of $35.35534 \, \mathrm{kpc} \, \mathrm{Gyr}^{-1}$.
    The particles are evolved with a cross-section of $\sigma_0/m_\chi = 7 \times 10^7 \, \mathrm{cm}^2 \, \mathrm{g}^{-1}$, $w = 10.0 \, \mathrm{km} \, \mathrm{s}^{-1}$ and the SIDM kernel sizes are computed using $N_\mathrm{ngb} = 64$.
    }
    \label{fig:gadget_test_cosmo}
\end{figure}

For testing the implementation of velocity-dependent self-interactions, we introduce and use a new test problem in this Appendix.
The test problem is very similar to the one we have used by \cite{Fischer_2022}.
A single particle is travelling within an expanding space through a background density.
This background is at rest (zero canonical momentum) and has no density gradient.
For this test problem, we only consider the drag force which decelerates the particle but do not re-add the energy as described in section~2.2 of \cite{Fischer_2021a}.
We do not take any further physics into account, i.e.\ run the test problem without gravity.
Hence, we expect the test particle to be decelerated over time.
We calculate semi-analytically how the canonical momentum of the test particle evolves over time and compare the results from the simulation to it.
This is shown in Fig.~\ref{fig:gadget_test_cosmo}.
Note, in the absence of self-interactions the canonical
momentum would stay constant over the cosmic expansion.

Further, we want to point out that this test problem is more susceptible to numerical errors than a typically fSIDM simulation.
The interaction between a pair of numerical particles does not change their relative velocity.
This makes the pairwise interaction in some sense time-implicit and more stable.
However, if we compute the drag force only and do not re-add the energy, as done for the test problem, we break this.
This is true for the test problem when conducted with a velocity-independent cross-section \citep[see appendix A by][]{Fischer_2022} too.
But in contrast, the velocity dependence makes it even more unstable.
Assuming that the test particle is slightly faster than it is supposed to be, one would expect the drag force to be stronger (velocity-independent cross-section) or weaker (velocity-dependent cross-section) than it is supposed to experience.
The first case would suppress the deviation but the second enhances it.
In the opposite case where the particle is slower than supposed, one finds again that the velocity-dependent cross-section tends to increase the deviation.
In consequence, the test problem we show is quite unstable.
However, in general, this depends on how strong the velocity dependence is, in our model specified by $\alpha$.

Overall, we find that the test simulation agrees sufficiently enough with the prediction and we can conclude the implementation of velocity-dependent self-interactions works as supposed.

\section{Convergence of density profile} \label{sec:convergence_test}

\begin{table}
    \centering
    \begin{tabular}{c|c|c|c|c}
        name & $N_\mathrm{high\,res}$ & $m_\mathrm{DM}$ \\
        & & $[\mathrm{M_\odot} \, h^{-1}]$ \\
         \hline
        1x & $\sim4.51 \times 10^4$ & $8.3 \times 10^8$ \\
        10x & $\sim4.52 \times 10^5$ & $8.3 \times 10^7$ \\
        25x & $\sim1.13 \times 10^6$ & $3.3 \times 10^7$ \\
        250x & $\sim1.13 \times 10^7$ & $3.3 \times 10^6$ \\
        2500x & $\sim1.13 \times 10^8$ & $3.3 \times 10^5$
    \end{tabular}
    \caption{The properties of the zoom-in simulations we use for the convergence test are given.
    We provide the name of the simulation, the number of particles in the highly resolved region ($N_\mathrm{high\,res}$) and the mass of the high-resolution particles ($m_\mathrm{DM}$).
    All simulations share the same initial conditions but with different resolutions.}
    \label{tab:sim_props_zoom}
\end{table}

\begin{figure*}
    \centering
    \includegraphics[width=\textwidth]{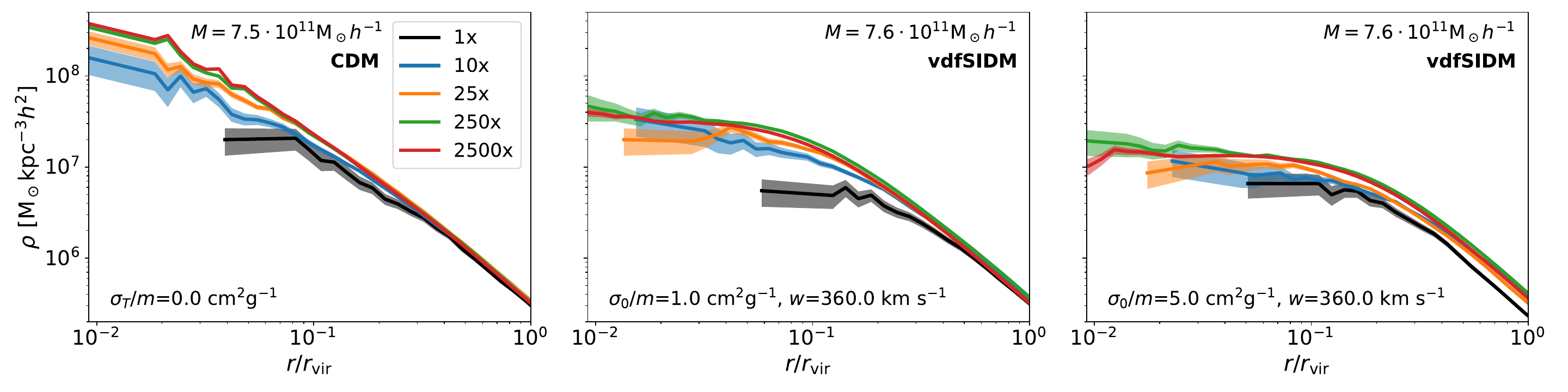}
    \caption{We show the density profile of the most massive subhalo of our zoom-in simulations.
    We give the profile for a CDM simulation (left-hand panel), which is the same as given by \citet{Fischer_2022}.
    The other two panels show the result of velocity-dependent fSIDM simulations.
    The different colours indicate different resolutions.
    This allows us to see that the profiles are converging for increasing resolution.
    In the highest resolved run, the subhalo is represented by $\approx 2.3 \times 10^6$ particles.}
    \label{fig:convergence}
\end{figure*}

In this Appendix, we study the convergence of our simulations with velocity-dependent self-interactions.
To do so we run zoom-in simulations of the same object but with varying resolutions.
The zoom-in region is selected from a large box with a comoving side length of $1 \, \mathrm{Gpc}\,h^{-1}$ and its most massive halo has a virial mass of $\sim 8.8 \times 10^{11} \, \mathrm{M_\odot} \, h^{-1}$.
Several publications \citep[e.g.][]{Planelles_2014, Rasia_2015} made use of this box for zoom-in initial conditions and it was first described by \cite{Bonafede_2011}.
We run the simulation with two cross-sections, $\sigma_0/m = 1.0 \, \mathrm{cm}^2 \, \mathrm{g}^{-1}$, $w = 360.0 \, \mathrm{km} \, \mathrm{s}^{-1}$ and $\sigma_0/m = 5.0 \, \mathrm{cm}^2 \, \mathrm{g}^{-1}$, $w = 360.0 \, \mathrm{km} \, \mathrm{s}^{-1}$.
The different resolutions we simulated are described in Tab.~\ref{tab:sim_props_zoom}.

In Fig.~\ref{fig:convergence}, we show the density profile for the most massive halo of the zoom-in simulations.
We can see that the density profile converges for collisionless DM (upper panel), but also when velocity-dependent self-interactions are present (middle and lower panel).

\section{Central density gradient of satellites} \label{sec:density_gradient}

\begin{figure*}
    \centering
    \includegraphics[width=\textwidth]{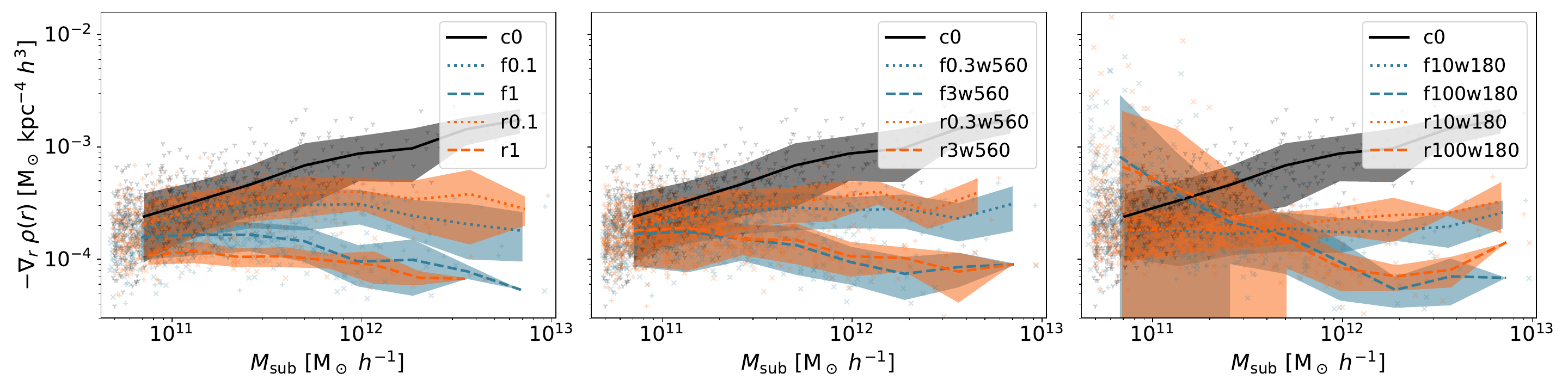}
    \caption{The central density gradient is shown for satellites with a mass of at least $\approx 4.9 \times 10^{10}\, \mathrm{M_\odot} \, h^{-1}$. We consider all satellites that are not the primary subhalo.
    The lines indicate the mean and the shaded regions the standard deviation for the corresponding DM models.
    This is analogous to Fig.~\ref{fig:vel_circ}, as well as the markers.}
    \label{fig:density_gradient}
\end{figure*}

In Fig.~\ref{fig:density_gradient}, we show the density gradient in the centre of satellites more massive than $\approx 4.9 \times 10^{10}\, \mathrm{M_\odot} \, h^{-1}$ as a function of their mass.
We consider all subhaloes identified by \textsc{SubFind} satellites if they are not a primary subhalo (see Sec.~\ref{sec:simulations}).
Note this figure is built analogously to Fig.~\ref{fig:vel_circ}.
The density gradient is computed using the mean density of the innermost 200 particles and the corresponding radius compared to the radius within which the average density drops by $50\%$. 

Similar to the circular velocity shown in Fig.~\ref{fig:vel_circ}, we find that for the strongly velocity-dependent cross-section (the right-hand panel of Fig.~\ref{fig:density_gradient}) the density gradient is on average steeper than for CDM at small satellite masses when the cross-section is sufficiently large. This indicates that the corresponding satellites are collapsing.
In contrast, we do not find these steep density gradients for the other cross-section with no (left-hand panel) or a weaker (middle panel) velocity dependence.
Moreover, the simulations for those cross-sections show density gradients that are on average flatter compared to CDM, i.e.\ those satellites host a density core.

\section{Frequent versus rare self-interactions} \label{sec:results_frequent_vs_rare_add}

\begin{figure*}
    \centering
    \includegraphics[width=0.65\columnwidth]{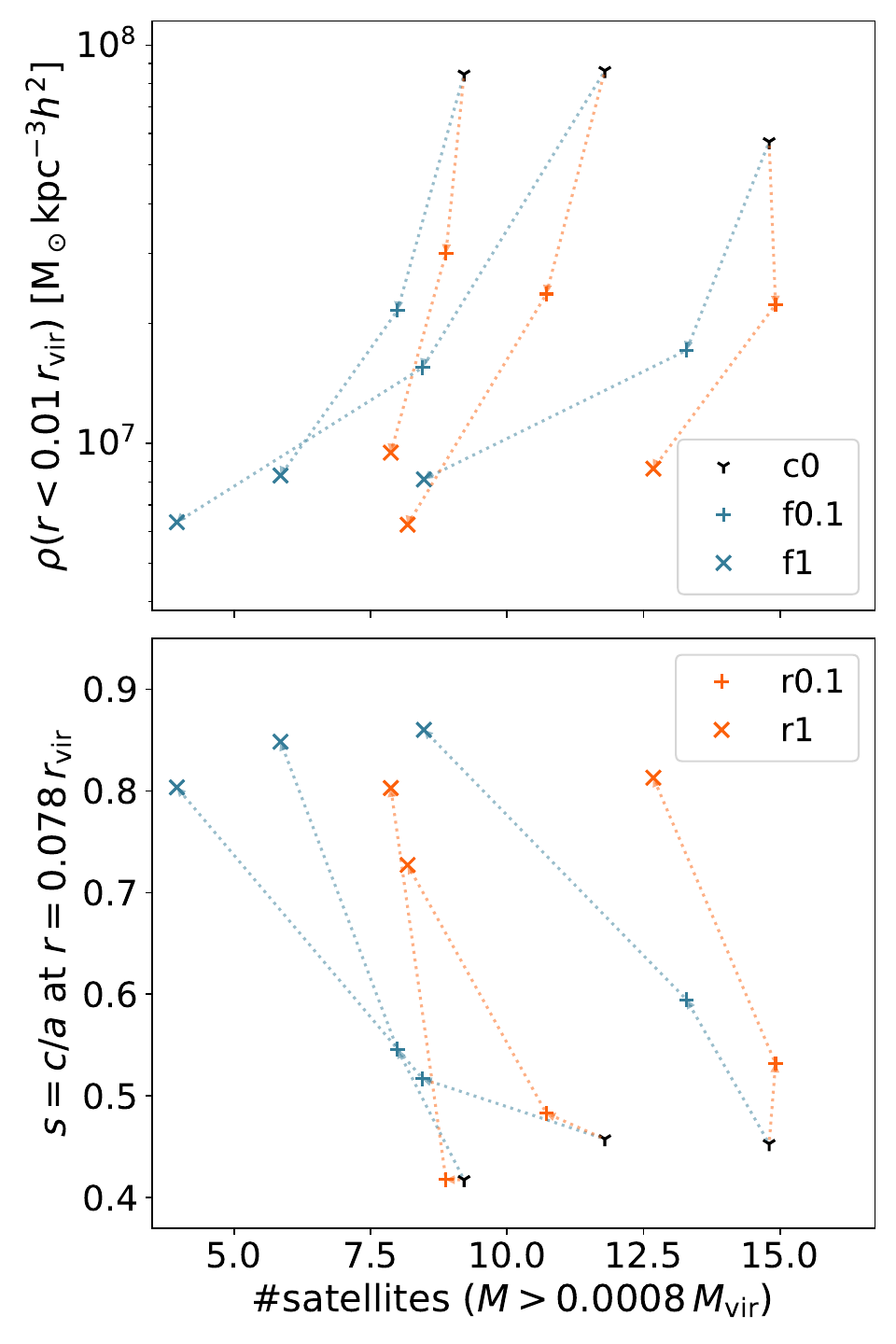}
    \includegraphics[width=0.65\columnwidth]{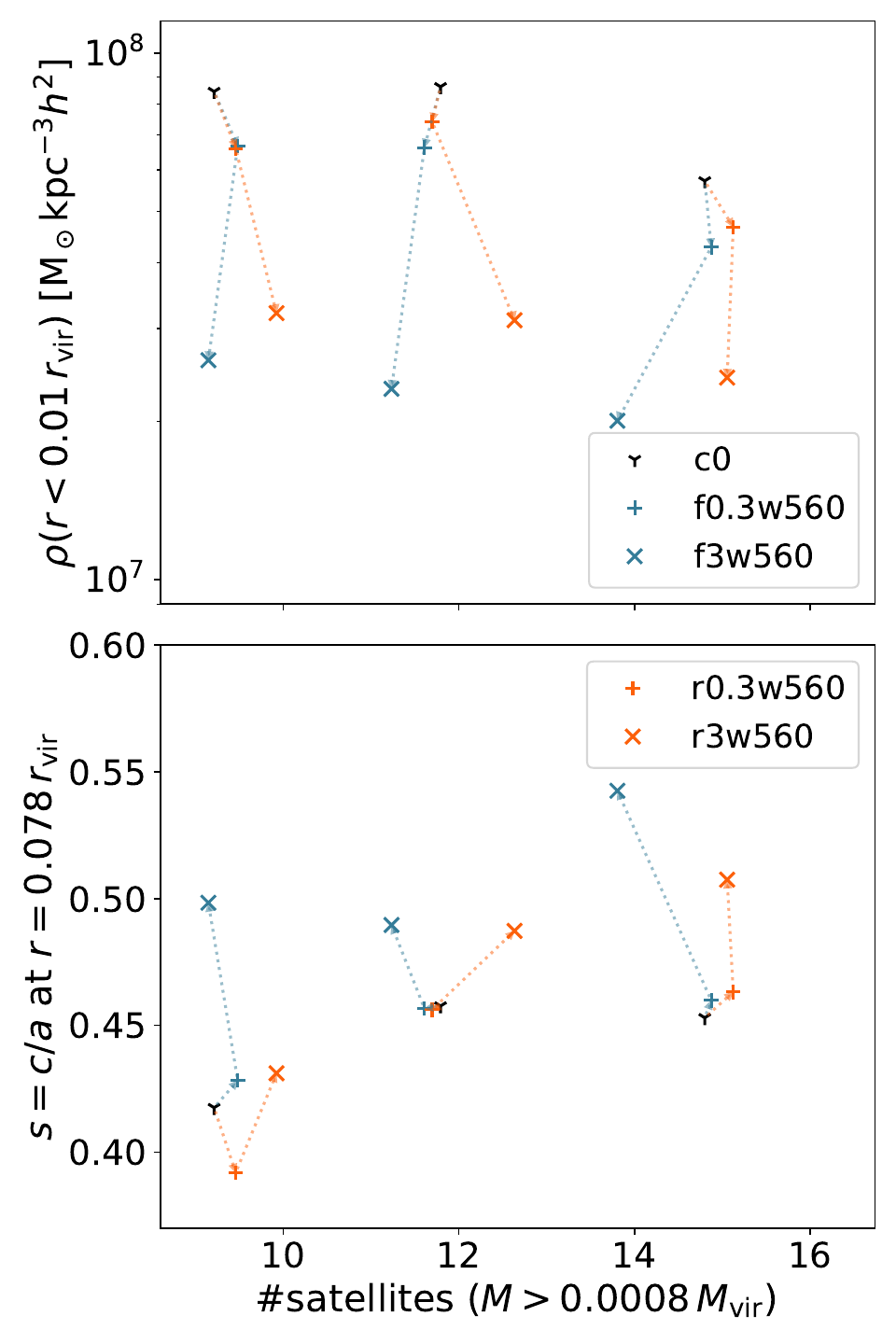}
    \includegraphics[width=0.65\columnwidth]{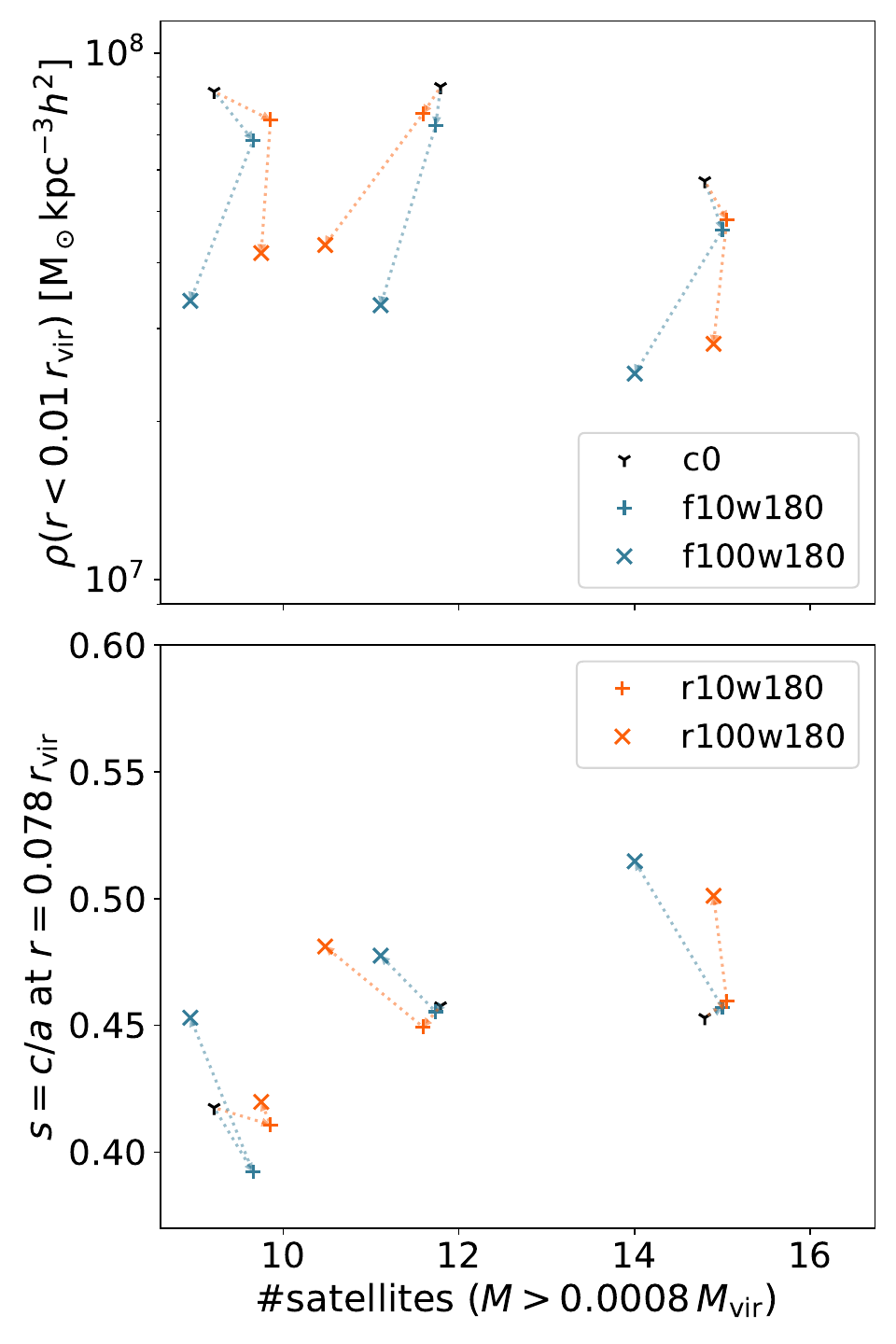}
    \caption{We show the central density (upper panels) and the shape (lower panels) of the host halo as a function of the number of their satellites for different DM models. The velocity-independent cross-sections are shown in the left-hand panels \citep[this has previously been shown in fig.~16 by][]{Fischer_2022}. The middle panels give the results for the models with $w=560 \, \mathrm{km} \, \mathrm{s}^{-1}$. And cross-sections with the strongest velocity dependence ($w=180 \, \mathrm{km} \, \mathrm{s}^{-1}$) are displayed in the right-hand side panels.}
    \label{fig:frequent_vs_rare_add}
\end{figure*}

Here, we show the central density of the host halo as a function of the number of satellites (Fig.~\ref{fig:frequent_vs_rare_add}).
The three most massive haloes are displayed as previously done in fig.~16 by \cite{Fischer_2022}.
We find that the frequent self-interactions independent of $w$ reduce the number of satellites stronger than rare scattering when comparing them at levels of the same central host density (upper panels) or the roundness of the host's shape (lower panels).
However, one halo for the $w=180 \, \mathrm{km} \, \mathrm{s}^{-1}$ poses an exception.

\section{SIDM constraints} \label{sec:constraints}

\begin{figure}
    \centering
    \includegraphics[width=\columnwidth]{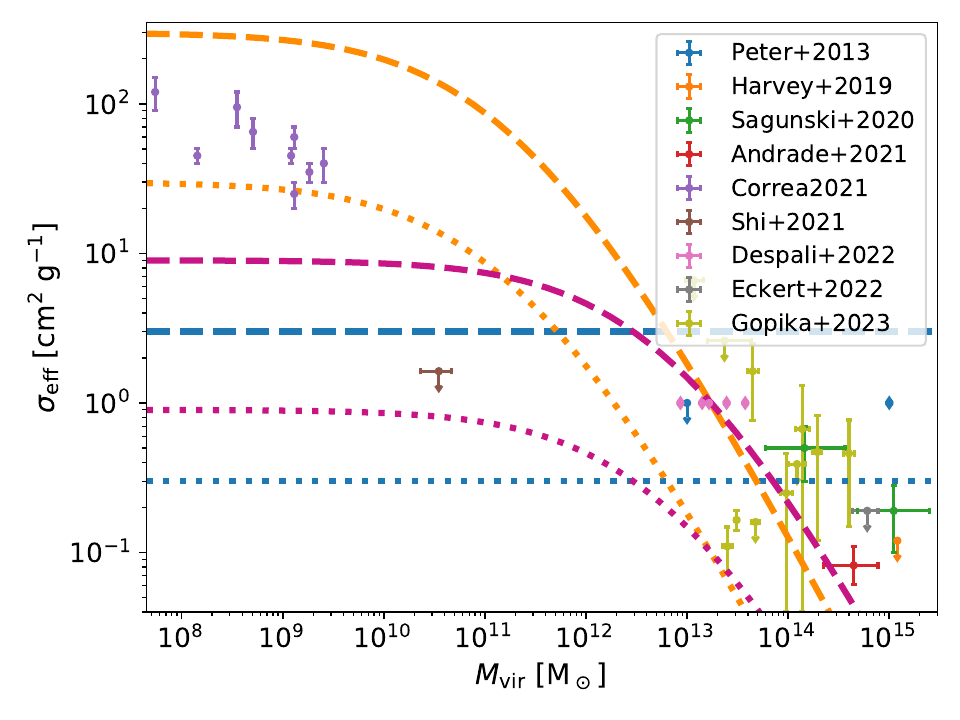}
    \caption{We show constraints for a velocity-independent cross-section together with the fSIDM models that we simulated. This is given in terms of the effective cross-section, $\sigma_\mathrm{eff}$ \citep{Yang_2022D} as a function of the virial DM halo mass. Constraints on the self-interaction strength obtained by various authors are shown. The colours of our SIDM models correspond to the ones shown in Fig.~\ref{fig:cross-sections}. Note, our rSIDM models are $1/3$ weaker than the fSIDM cross-sections when compared in terms of $\sigma_\mathrm{eff}$.}
    \label{fig:constraints}
\end{figure}

In Fig.~\ref{fig:constraints}, we show constraints on the strength of DM self-interactions together with our SIDM models.
Here, we compute the effective cross-section as introduced by \cite{Yang_2022D} (see also equation~\ref{eq:sigma_eff}).
This requires an estimate of an effective velocity dispersion, which we compute from a given virial mass, $M_\mathrm{vir}$.
To do so, we use the halo mass-concentration relation given by \cite{Dutton_2014}.
With the obtained concentration parameter, $c$, we infer the maximum velocity dispersion, $\nu^2_\mathrm{max}$.
For the effective velocity dispersion we employ $\sigma^\mathrm{eff}_\mathrm{1D} = 0.9 \times \nu_\mathrm{max}$.
We choose the factor of $0.9$ as it provides a good match for our isolated NFW simulation shown in Section~\ref{sec:results_nfw}.
This concerns the match of the velocity-independent cross-sections with the ones that are described by $w=720 \, \mathrm{km}\,\mathrm{s}^{-1}$ and $\sigma_0 / m = 5 \times 10^3 \, \mathrm{cm}^2 \, \mathrm{g}^{-1}$.
However, we have to note that the viscosity cross-section-like matching for the angular dependence in $\sigma_\mathrm{eff}$ does not provide a match as good as the one from the momentum transfer cross-section in this particular case. If we would have used the viscosity cross-section for the matching the isotropic cross-section would have $3/2$ of the strength we obtained from the momentum transfer matching while leaving the fSIDM cross-section unchanged.

The constraints shown in Fig.~\ref{fig:constraints} stem from measures of different effects that SIDM has on the distribution of DM. This includes the formation of a density core \citep{Sagunski_2021, Andrade_2021, Correa_2021, Shi_2021, Eckert_2022, Gopika_2023}, oscillations of the brightest cluster galaxy \citep{Harvey_2019}, and the shapes of the haloes \citep{Peter_2013, Despali_2022}.


\bsp	
\label{lastpage}
\end{document}